\documentclass[a4paper,twocolumn,10pt,final,accepted=2020-05-21]{quantumarticle}
\pdfoutput=1
\usepackage[utf8]{inputenc}
\usepackage[english]{babel}
\usepackage[T1]{fontenc}
\usepackage{amsmath,amssymb}
\usepackage[colorlinks=true,breaklinks=true,allcolors=blue]{hyperref}
\usepackage[numbers,sort&compress]{natbib}

\usepackage{bbm}
\usepackage{mathrsfs} 
\usepackage{comment}

\allowdisplaybreaks

\newcommand{\lla}{\left\langle}
\newcommand{\rra}{\right\rangle}
\newcommand{\pt}{\partial}

\newcommand{\di}{\,\mr{d}}
\newcommand{\mr}[1]{\mathrm{#1}}

\newcommand{\1}{\mathbbm 1}

\newcommand{\br}[1]{\langle #1 |} 
\newcommand{\kt}[1]{| #1 \rangle} 
\newcommand{\brkt}[2]{\langle #1 | #2 \rangle} 
\newcommand{\bropkt}[3]{\langle #1 |#2| #3 \rangle} 
\DeclareMathOperator{\diam}{\text{diam}}

\renewcommand{\d}{\delta}

\newcommand{\s}{\sigma}

\newcommand{\D}{\Delta}

\newcommand{\kp}{\kappa}

\begin{document}
		

\title{Modelling equilibration of local many-body quantum systems by random graph ensembles}
\author{Daniel Nickelsen}
\email[]{danielnickelsen@sun.ac.za}
\orcid{0000-0001-8052-9286}
\affiliation{National Institute for Theoretical Physics (NITheP), Stellenbosch 7600, South Africa}
\affiliation{Institute of Theoretical Physics,  Department of Physics, University of Stellenbosch, Stellenbosch 7600, South Africa}
\affiliation{African Institute for Mathematical Sciences, Muizenberg, Cape Town, South Africa}
\author{Michael Kastner}
\email[]{kastner@sun.ac.za}
\orcid{0000-0002-6787-0065}
\affiliation{National Institute for Theoretical Physics (NITheP), Stellenbosch 7600, South Africa}
\affiliation{Institute of Theoretical Physics,  Department of Physics, University of Stellenbosch, Stellenbosch 7600, South Africa}


\begin{abstract}
We introduce structured random matrix ensembles, constructed to model many-body quantum systems with local interactions. These ensembles are employed to study equilibration of isolated many-body quantum systems, showing that rather complex matrix structures, well beyond Wig\-ner's full or banded random matrices, are required to faithfully model equilibration times. Viewing the random matrices as connectivities of graphs, we analyse the resulting network of classical oscillators in Hilbert space with tools from network theory. One of these tools, called the maximum flow value, is found to be an excellent proxy for equilibration times. Since maximum flow values are less expensive to compute, they give access to approximate equilibration times for system sizes beyond those accessible by exact diagonalisation.
\end{abstract}

\maketitle

\section{Introduction}
\label{sec:intro}
By means of typicality arguments, equilibration of isolated quantum systems has been rigorously proved to occur under rather general conditions \cite{vonNeumann29,Tasaki98,Linden_etal09,Goldstein_etal10,Reimann10,KastnerReimann12,GogolinEisert16,Mori_etal18}. It remains however a key open problem to identify which properties of quantum systems are responsible for ensuring that equilibration timescales are physically realistic \cite{ShortFarrelly12,GoldsteinHaraTasaki13,GoldsteinHaraTasaki14,Malabarba_etal14,Farrelly2016,Reimann16,Wilming_etal,GarciaPintos_etal17,deOliveira_etal18,Reimann2019}. As in virtually all other problems of many-body quantum physics, the exponential growth of Hilbert space dimension with system size renders direct studies infeasible already for moderate system sizes. To circumvent this difficulty in models of nuclei of heavy atoms, Wigner proposed the use of random matrices as Hamiltonians that statistically share certain properties of the nuclei \cite{Wigner1955,Wigner1957,Wigner1958}. One such choice, the so-called Gaussian ensembles, is given by matrices where all elements are normally distributed i.i.d.\ numbers normalized to some measure \cite{Mehta2004}. Gaussian random matrices are a particularly convenient choice, as they allow for the analytic calculation of ensemble-averaged properties like limiting distributions of eigenvalue statistics, i.e.\ Wigner's semi-circle law \cite{Wigner1958}. Also, the level-spacing statistics of Gaussian random matrices are known to follow the Wigner-Dyson distribution \cite{Mehta2004,GogolinEisert16}, which implies that small spacings between eigenvalues of the Hamiltonian are rare. 

Gaussian ensembles of random matrices have also been used to study equilibration and thermalisation of isolated quantum systems \cite{Santos2017,Torres-Herrera2018,Santos2018}. Since the time evolution of expectation values of observables is governed by differences of energy eigenvalues, the level-spacing statistics of random matrices influences the equilibration times \cite{Torres-Herrera2016}. The suppression of small spacings, as in the Wigner-Dyson distribution, favours short equilibration times. Indeed, in Ref.~\cite{GoldsteinHaraTasaki14} it has been shown that equilibration under random-matrix Hamiltonians typically happens extremely fast, much faster than everyday experience or the analysis of models from condensed matter theory suggests. This finding suggests that Gaussian random matrices are unsuitable for representing physically realistic quantum many-body Hamiltonians.

Locality, in the sense of short-ranged interactions between only few constituents in a spatially extended many-body system, is believed to be a key requirement for realistic equilibration dynamics. Gaussian random matrices typically represent fully-connected systems with distance-independent interaction strengths between constituents. To incorporate locality into random matrix ensembles, early works in nuclear physics have introduced the two-body random ensemble \cite{French1970,Bohigas1971}. This ensemble uses Gaussian random matrices to model the two-body interactions in a fermionic systems, and assembles these matrices into the total Hamiltonian of a many-body system, which turns out to be a sparse matrix. The two-body random ensemble has also been used to study thermalisation of isolated fermionic systems \cite{Kota2011,FlambaumIzrailev01,BorgonoviIzrailevSantos19,BorgonoviIzrailev19}. Generalisations of the two-body random ensemble to random $k$-body interactions have been studied for fermionic as well as bosonic systems \cite{Kota2001,Kota2018,Vyas2018}. More generally, random matrix ensembles that use full random matrices for local interactions to assemble random total Hamiltonians are referred to as {\em embedded ensembles}.

If one examines the structure of matrices from the embedded ensembles, one finds that they are sparse matrices and that they display a banded structure. Similarly, the (non-random) Hamiltonian of a many-body system with local interactions, when represented in the tensor product basis with respect to which locality is defined, turns out to be a banded, sparse matrix. In fact, already Wigner in his early attempts to model nuclei of heavy atoms introduced banded matrices that have nonzero elements only up to a certain distance from the diagonal \cite{Wigner1955,Wigner1957}. In addition to resembling Hamiltonians with local interactions, several studies have shown that random band matrices are also capable of modelling certain aspects important for equilibration timescales, e.g. the transition from Wigner-Dyson statistics for chaotic system to Poisson statistics of regular systems, and the emergence of many-body localisation \cite{Seligman1985,Fyodorov1996,Erdos2011}. Variants of Wigner band matrices include power-law banded random matrices \cite{Mirlin1996,Nosov2019} where the standard deviation of matrix elements decreases with the distance from the main diagonal according to a power law, and sparse random band matrices \cite{Mendez-Bermudez2017}, which also account for the possibility of having vanishing elements in the band. However, while the mentioned ensembles of banded random matrices may model certain aspects of physical systems more realistically, they come with the disadvantage that analytic calculations of their spectral properties are far more intricate \cite{Jana2017,Bourgade2018,Dumitriu2019}. An overview of applications of embedded and banded ensembles is provided in Ref.~\cite{Borgonovi2016}. In a different attempt to include aspects of locality into random Hamiltonians, the authors of Ref.~\cite{Brandino2012} disregarded the banded structure of local Hamiltonians, and instead investigated only the effect of the sparseness of the matrix on the equilibration dynamics.

In this paper we address in a systematic way the interconnection between the locality of a many-body Hamiltonian and the structural properties (bandedness, sparseness, etc.)\ of its matrix representation. To this purpose, we interpret the Hamiltonian matrix as the connectivity matrix of a weighted graph, which allows us to apply concepts from graph and network theory. As our first main result, we identify key features of the graphs corresponding to local Hamiltonians, including degree distributions as well as a node-dependent (``curved'') bandedness of the connectivity matrix. As a second main result we identify the matrix (or graph) properties that are essential to faithfully reproduce the typical equilibration times of a local many-body system. We find that a random Hamiltonian with the correctly curved bandshape and the correct, constant degree faithfully reproduces the equilibration times of a local system. Abandoning either the correct bandshape or constant degree, however, spoils the success. We also demonstrate that Hamiltonians from such ensembles can be constructed efficiently. Our third main result is that the maximum flow, a tool from network theory, can serve as a proxy for equilibration times. This means that, instead of computing the equilibration time of a Hamiltonian, it is sufficient---and numerically more efficient---to calculate its maximum flow value, from which the equilibration time can be deduced. Combining the efficient graph construction and the calculation of the maximum flow value, equilibration times of local Hamiltonians can be investigated more efficiently.

\section{Equilibration}
\label{sec:set}
We choose the equilibration dynamics of a quantum many-body system as our diagnostic tool with which we analyse whether the locality of interactions is successfully modelled by a random matrix ensemble. Defining equilibration in an isolated, hence unitarily evolving, quantum system requires some care, as we discuss in the following. We consider the Schr\"odinger time evolution
\begin{equation}\label{eq:SE}
i\partial_t\kt{\psi(t)}=H\kt{\psi(t)},
\end{equation}
generated by a time-independent Hamiltonian $H$. Under the dynamics of \eqref{eq:SE}, the infinite time average of the expectation value of an observable $O$ can be written as
\begin{multline}\label{eq:longtime}
\lim_{T\to\infty} \frac{1}{T}\int_0^T \bropkt{\psi(t)}{O}{\psi(t)} \di t \\
= \sum_j |c_j|^2 \bropkt{e_j}{O}{e_j} \equiv \lla O \rra_\text{diag},
\end{multline}
where $\kt{e_j}$ are energy eigenstates and $c_j=\brkt{e_j}{\psi_0}$ are the overlaps with the initial state. Unlike the microcanonical ensemble that uses equal weights, the so-called {\em diagonal ensemble}\/ average in the second line of \eqref{eq:longtime} uses the weights $|c_j|^2$. If the system equilibrates, it is reasonable to expect that the expectation value of the observable $O$ equilibrates to the value $\lla O \rra_\text{diag}$. Through the dependence of the coefficients $c_j$ on $\psi_0$, the equilibrium value will in general depend on the initial state. 

In any system with a finite-dimensional Hilbert space, unitary time evolution implies that $\lla O \rra$ evolves quasi-periodically in time. As a consequence, even at arbitrarily late times, fluctuations around the equilibrium value $\lla O \rra_\text{diag}$ occur. Because of these fluctuations, a sensible definition of equilibration in isolated macroscopic quantum systems requires a probabilistic perspective. The largest part of the Hilbert space, referred to as the {\em equilibrium subspace}, is spanned by eigenstates $\kt{o_j}$ of $O$ that have eigenvalues $o_j$ close to the equilibrium value, $o_j\approx\lla O \rra_\text{diag}$. The complement of this subspace is the {\em nonequilibrium subspace}, spanned by eigenstates of $O$ with eigenvalues that appreciably differ from $\lla O \rra_\text{diag}$. Equilibrium states are then defined as states that have a large overlap with the equilibrium subspace; see \cite{GoldsteinCommentary10,GoldsteinHaraTasaki14} for details. As a consequence of this definition, an equilibrium state $\kt{\psi}$ not only guarantees that $\bropkt{\psi}{O}{\psi}\approx \lla O \rra_\text{diag}$, but also that fluctuations around this expectation value are small. For a many-body system with large Hilbert space dimension, the dimension of the equilibrium subspace is typically much larger than the dimension of the nonequilibrium subspace. As a consequence, a state drawn randomly from the Hilbert space according to the Haar measure will most likely be an equilibrium state \cite{GoldsteinCommentary10,GoldsteinHaraTasaki14}. Moreover, a nonequilibrium initial state typically evolves towards equilibrium and stays there for most times \cite{Linden_etal09,Goldstein_etal10}. Sporadic recurrences to nonequilibrium occur, but with a probability that becomes vanishingly small with increasing system size.

Throughout this section, equilibration was discussed with reference to a specific observable $O$, and also the (non)equilibrium subspaces have been defined with respect to that observable. The reason for this choice is related to the quasi-periodic behaviour of a unitarily evolving quantum system on a finite-dimensional Hilbert space. In such a system it is always possible to find observables (i.e., Hermitian operators) that either do not evolve at all (like the projectors $\kt{e_j}\br{e_j}$ onto energy eigenspaces), or that evolve periodically and hence do not equilibrate (like $\kt{e_i}\br{e_j}+\kt{e_j}\br{e_i}$). The physical relevance of these and similar observables is, however, questionable, as they are nonlocal quantities (as defined in Sec.~\ref{sec:spin}) and usually inaccessible in experimental measurements of many-body systems. Defining equilibration with respect to specific observables is hence not a shortcoming, but a desired feature.

\section{Network picture of equilibration}
\label{sec:graph}
The fact that equilibration in Sec.~\ref{sec:set} has been defined with respect to a specific observable suggests that, for the physical problem at hand, the eigenbasis $\kt{o_j}$ of $O$ may play a distinguished role. In this basis, the Schr\"odinger equation \eqref{eq:SE} reads
\begin{align} \label{eq:SE2}
\dot x_j(t) = -i\sum_k H_{jk}\,x_k(t),
\end{align}
where
\begin{align}\label{eq:x}
x_j(t)&=\brkt{o_j}{\psi(t)},& H_{jk}&=\bropkt{o_j}{H}{o_k}.
\end{align}
The eigenstates $\kt{o_j}$ are assumed to be sorted in ascending order, $o_{j+1}\geq o_j$. By componentwise integration, Eq.~\eqref{eq:SE2} can be written as
\begin{align} \label{eq:netequ}
e^{iH_{jj}t}x_j(t) = x_j(0) - i\sum_{k\neq j} H_{jk} \int_0^t e^{iH_{jj}\tau} x_k(\tau) \di\tau. 
\end{align}
Taking the diagonal entries $H_{jj}$ as frequencies and the off-diagonal entries $H_{jk}$ as coupling strengths between classical oscillators, we can interpret this equation as the description of a network of coupled oscillators with amplitudes $x_j(t)$. These observations motivate to use graph-theoretical tools for the analysis of the equilibration dynamics.

Using this idea of coupled oscillators, the network picture of equilibration can be described as follows \cite{Nickelsen2019}. Preparing the system in an initial state where oscillators of equilibrium states have vanishing amplitudes $x_j(t)$, oscillations need to propagate through the network in order to activate the equilibrium oscillators. Once these oscillators are active, measurement probabilities $x_j$ for equilibrium values $o_j\approx \lla O\rra_\text{diag}$ are dominating and, according to the definition of equilibrium in Sec.~\ref{sec:set}, the system has equilibrated. The efficiency or speed of equilibration is expected to depend on the network properties like degree distribution, connectivity, etc. A more detailed motivation of such a network picture of equilibration is given in Appendix \ref{app:netequi}. The main goal of the present paper is to analyse the properties of such networks for physically realistic quantum many-body systems and link those network properties to the equilibration behaviour. 

We consider a weighted graph consisting of $N$ nodes, where $N$ is the dimension of the Hilbert space of the quantum system. To each node $j$ we assign the classical complex degree of freedom $x_j$ defined in \eqref{eq:x}. The weights of the graph edges are given by $|H_{jk}|$. The graph topology is described by the adjacency matrix $A$ with matrix elements
\begin{equation}\label{eq:adjacency}
A_{jk}=
\begin{cases}
0 & \text{if $H_{jk}=0$},\\
1 & \text{else}.
\end{cases}
\end{equation}
We use
\begin{equation}
d_O(j):=\bigl|o_j-\lla O\rra_\text{diag}\bigr|
\end{equation}
as a distance of the nodes to equilibrium, which will serve as a metric on the graph: nodes $j$ with $d_O(j)\approx0$ are equilibrium nodes, and those with $d_O(j)\not\approx 0$ are nonequilibrium nodes.

In Ref.~\cite{Nickelsen2019}, this network picture of quantum dynamics has been used to derive a lower bound on the timescale at which the system equilibrates. Remarkably, the bound was found to depend on the degrees of locality of the Hamiltonian $H$ as well as of the observable $O$. There exist various notions of locality, and we will introduce two of them in Sec.~\ref{sec:spin}, but we expect all of them to have in common that the corresponding network or graph is less strongly connected when both $H$ and $O$ are local. Weaker and/or fewer connections are then expected to lead to longer equilibration timescales, and this is indeed what was found in Ref.~\cite{Nickelsen2019}.

There exists a zoo of measures of graph topology, and our aim is to identify a measure that can serve as a proxy for equilibration timescales. In view of the above described intuitive picture of an oscillation of a nonequilibrium node propagating through the network to an equilibrium node along all of the existing pathways between the nodes, promising candidates are measures that take into account multiple pathways. The node connectivity, defined as the minimum number of nodes that need to be removed in order to split the graph into two disconnected sub-graphs, takes multiple paths into consideration. A closely related graph measure is the maximum flow value $f_\text{max}$, for which the weights $|H_{jk}|$ of the graph are taken as capacities rendering the graph a flow network, where $f_\text{max}$ is the total capacity between two nodes. 
%
%
Visualizing the flow as a fluid entering the network at node $k$ and leaving it at node $j$, $f_\text{max}(k,j)$ is the maximum amount of fluid per unit time that can be routed through the network using those two nodes. 

Most algorithms use the max-flow min-cut theorem to calculate $f_\text{max}$. The theorem considers cuts of the network into two disconnected parts. A capacity value is assigned to each cut, which is the sum of all $|H_{uv}|$ that have been removed for the cut. The minimum cut is the one that is minimal in this capacity value. The max-flow min-cut theorem then states that $f_\text{max}(k,j)$ is identical to the capacity value of the minimum cut separating nodes $k$ and $j$. Intuitively, this means that the maximum flow that can be routed through the network is defined by the smallest bottleneck in capacity. A more formal definition of $f_\text{max}(k,j)$ can be found in Ref. \cite{Diestel2005}.
We will apply the maximum flow measure to the network picture of quantum spin models in Sec.~\ref{ssec:fmax}.

\section{Spin model}
\label{sec:spin}
There exist various notions of locality, and our goal is to explore network features and exploit graph measures for quantum many-body systems with different degrees of locality. To this aim, we introduce a rather general two-parameter family of quantum spin chains, where each of the two parameters $d$ and $n$ controls the degree of locality according to one of those notions of locality. The parameter $d$ quantifies the degree of locality according to the convention of, e.g., statistical physics or condensed matter physics, where nearest-neighbour interactions correspond to $d=1$, next-nearest-neighbour interactions to $d=2$, and so on. The parameter $n$ quantifies locality in the sense of the number of lattice sites involved in an interaction term, independently of the distance between the sites. A noninteracting Hamiltonian corresponds to $n=1$, a Hamiltonian consisting of pair interaction terms corresponds to $n=2$, etc.

Consider a quantum spin chain on a one-di\-men\-sion\-al lattice $\mathscr{L}$ consisting of $L$ sites. (Not to be confused with the nodes of the graph introduced in Sec.~\ref{sec:graph}.) For any subset $l\subseteq\mathscr{L}$ we denote by $\diam(l)$ the diameter of $l$, i.e., the largest Euclidean distance between any two sites in $l$. A  general Hamiltonian for a spin-$1/2$ chain on $\mathscr{L}$ with locality degrees $d$ and $n$ is given by
\begin{align} \label{eq:H}
H = \sum_{\d=1}^{d} \sum_{\chi(n)|\d} \sum_{\phi(n)} a(\chi,\phi) \prod_{i=1}^n \s_{\chi_i}^{\phi_i},
\end{align}
where $\chi(n)|\d$ denotes the set of all $n$-element subsets $l$ of $\mathscr{L}$ with the condition that $\diam(l)=\delta$. The set $\phi(n)$ consists of all $n$-tuples of orientations $x$, $y$, or $z$. $\s_i^{x}$, $\s_i^{y}$ and $\s_i^{z}$ denote the $x$, $y$ and $z$ components of Pauli spin operators acting on lattice site $i$. The coupling constants $a(\chi,\phi)$ are normal-distributed i.i.d.\ numbers with zero mean and standard deviation $1/2$. The random character of the couplings guarantees that accidental cancellations in the matrix elements $H_{jk}$ occur with zero probability, and hence the elements of the adjacency matrix \eqref{eq:adjacency} do not vanish by chance. Hamiltonians with parameter values $d=1$ and $n=2$ correspond to pair-interactions where each spin component of one spin interacts with all spin components of only its nearest neighbours. For $d=L-1$ and $n=L$ the Hamiltonian is the sum of products $\s_{1}^{\phi_1}\s_{2}^{\phi_2}\cdots\s_{L}^{\phi_L}$ over all sets $\phi$, in which case the Hamiltonian is maximally nonlocal. 

Our aim is to compare the equilibration timescales of Hamiltonians of the type \eqref{eq:H} for various values of the parameters $d$ and $n$. To make sure that such a comparison focusses on the degree of locality, we normalise all Hamiltonians such that $\|H\|=1$ to get rid of overall factors in $H$, as a Hamiltonian $2H$ would equilibrate twice as fast as $H$ for trivial reasons. Unnormalised Hamiltonians usually scale extensively or superextensively with the lattice size $L$, and our normalisation convention will therefore lead to longer equilibration times compared to the unnormalised counterpart. We will comment on implications of the normalisation at the end of Sec.~\ref{ssec:ens}.

The magnetisation is a natural observable in a quantum spin system. We study equilibration with respect to a weighted variant of the magnetisation in $z$-direction,
\begin{align}\label{eq:M}
M =  \frac{1}{L}\sum_{i=1}^L a_m(i)\,\s_i^{z},
\end{align}
with real weights $a_m$. The eigenstates $\kt{m_j}$ of $M$ are products of eigenstates of $\s_i^{z}$. For the case of homogeneous weights, $a_m(i)\equiv1$ for all $i$, the eigenvalues $m_i$ of $M$ take on values from the set $\{-L,L+2,\dots,L-2,L\}/L$. Most of these eigenvalues occur with high degeneracies, and hence the eigenstates $\kt{m_i}$ cannot be uniquely sorted according to their eigenvalues. To avoid this, we introduce small variations by choosing the $a_m$ as normal-distributed random numbers with mean $1$ and standard deviation $1/10$. A unique order of eigenstates brings out the graph properties more clearly and demonstrates that our analysis works for generic observables. Lifting the degeneracy, however, is not a requirement for our analysis, as is demonstrated in Appendix \ref{app:nonrandom}. Due to the independent, anisotropic and inhomogeneous coupling, we expect the equilibrium value $\lla M\rra_\text{diag}$ to be close to zero, which is confirmed by exact diagonalisation. Hence, equilibrium eigenstates are at the centre and nonequilibrium eigenstates at the edges of the spectrum.

\section{Results}\label{sec:res}

\subsection{Graph properties}\label{ssec:feat}

In this section we mostly study the properties of the (unweighted) graph defined through the adjacency matrix $A$ \eqref{eq:adjacency} that is associated with the Hamiltonian \eqref{eq:H} when represented in the eigenbasis of the magnetisation $M$ \eqref{eq:M}. We focus in particular on the dependence of the graph properties on the locality parameters $n$ and $d$ introduced in \eqref{eq:H}.

The degree distribution is a prominent graph property. The degree $\varrho(j) = \sum_{k\neq j} A_{jk}$ of a node $j$ counts the number of edges of this node. The degree distribution is the probability distribution of the degrees over all nodes $j$. Scale-free and small-world networks, for instance, have power-law distributed degrees, and the degrees of Bernoulli random graphs follow a Poisson distribution. Here we find that, for any given $n$ and $d$, all nodes of the graph have identical degrees. Such a graph is called a regular graph. By examining constant degrees $\varrho$ for system sizes up to $L=10$ and all possible combinations of locality parameters $n$ and $d$ (and up to $L=16$ for small values of $n$ and $d$), we conjecture that
\begin{align}\label{eq:degdist}
	\varrho(L,n,d) &=\sum_{q=0}^{n-1} \bigg[L- q\,\frac{d+1}{q+1}\bigg] \binom{d}{q}.
\end{align}
See Appendix \ref{app:degree} for details.

\begin{figure}[t]
	\includegraphics[width=\linewidth]{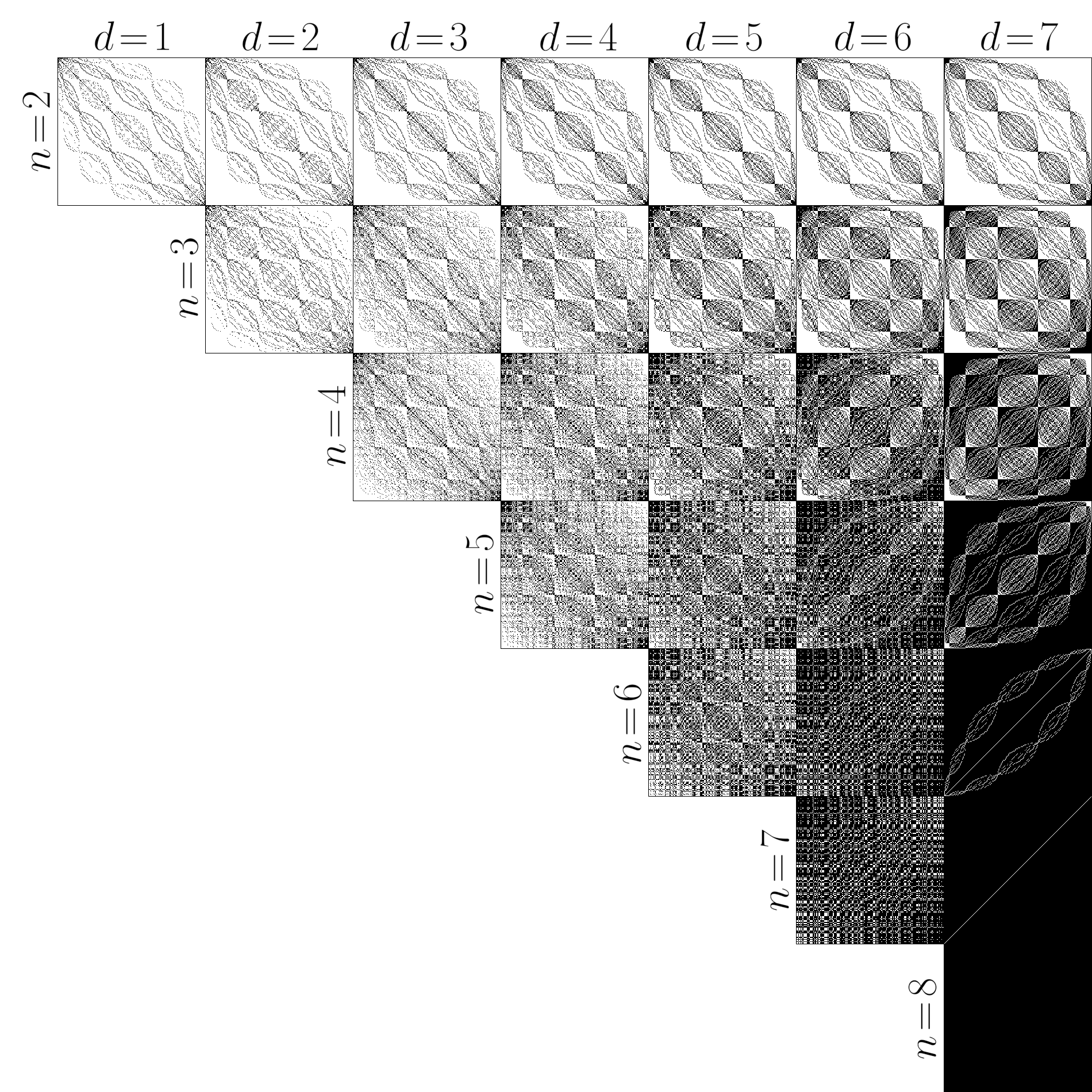}
	\caption{Plots of the adjacency matrices $A$ corresponding to the Hamiltonian \eqref{eq:H} in the eigenbasis of the magnetisation \eqref{eq:M} for a chain of $L=8$ spins and all possible combinations of values for $n$ and $d$. Black indicates nonzero matrix elements, white indicates zero elements.
	\label{fig:masks}}
\end{figure}

For a first impression of the effect of locality on the adjacency matrix $A$, Fig.~\ref{fig:masks} shows graphical representations of the matrix elements $A_{jk}$ for different choices of the parameters $n$ and $d$. The nearest-neighbour pair-interaction Hamiltonian [$n=2$ and $d=1$ in \eqref{eq:H}, top left in Fig.~\ref{fig:masks}] is a sparse matrix with a rather low density of nonzero elements. The maximally nonlocal Hamiltonian ($n=L$ and $d=L-1$, bottom right in Fig.~\ref{fig:masks}), on the other hand, is a dense matrix, which may be modelled by full random matrices of the Gaussian orthogonal ensemble. Other choices of $n$ and $d$ create matrices of intermediate sparsity.

The plots in Fig.~\ref{fig:masks} reveal that, in addition to being sparse, the adjacency matrices of sufficiently local systems are banded, in the sense that, outside a certain region around the main diagonal, all matrix elements are zero. Such a banded structure agrees with Wigner's intuition that banded random matrices may be more suitable to model physical systems \cite{Wigner1955,Wigner1957}.  Also, a banded structure is consistent with a recent theorem by Arad, Kuwahara, and Landau \cite{AradKuwaharaLandau16}, which affirms that a local observable represented in the eigenbasis of a local Hamiltonian is a banded matrix, possibly decorated with exponentially suppressed nonzero elements outside the band. The reasoning that led to this result can be reversed to show that a local Hamiltonian is banded in the eigenbasis of the observable, again with the possibility of exponentially small matrix elements outside the bands \cite{Nickelsen2019}. However, for the choice of Hamiltonian and observable considered in the present paper, we find that the matrix elements $H_{jk}$ are strictly zero outside the band.

\begin{figure*}[t]
	\includegraphics[width=0.45\linewidth]{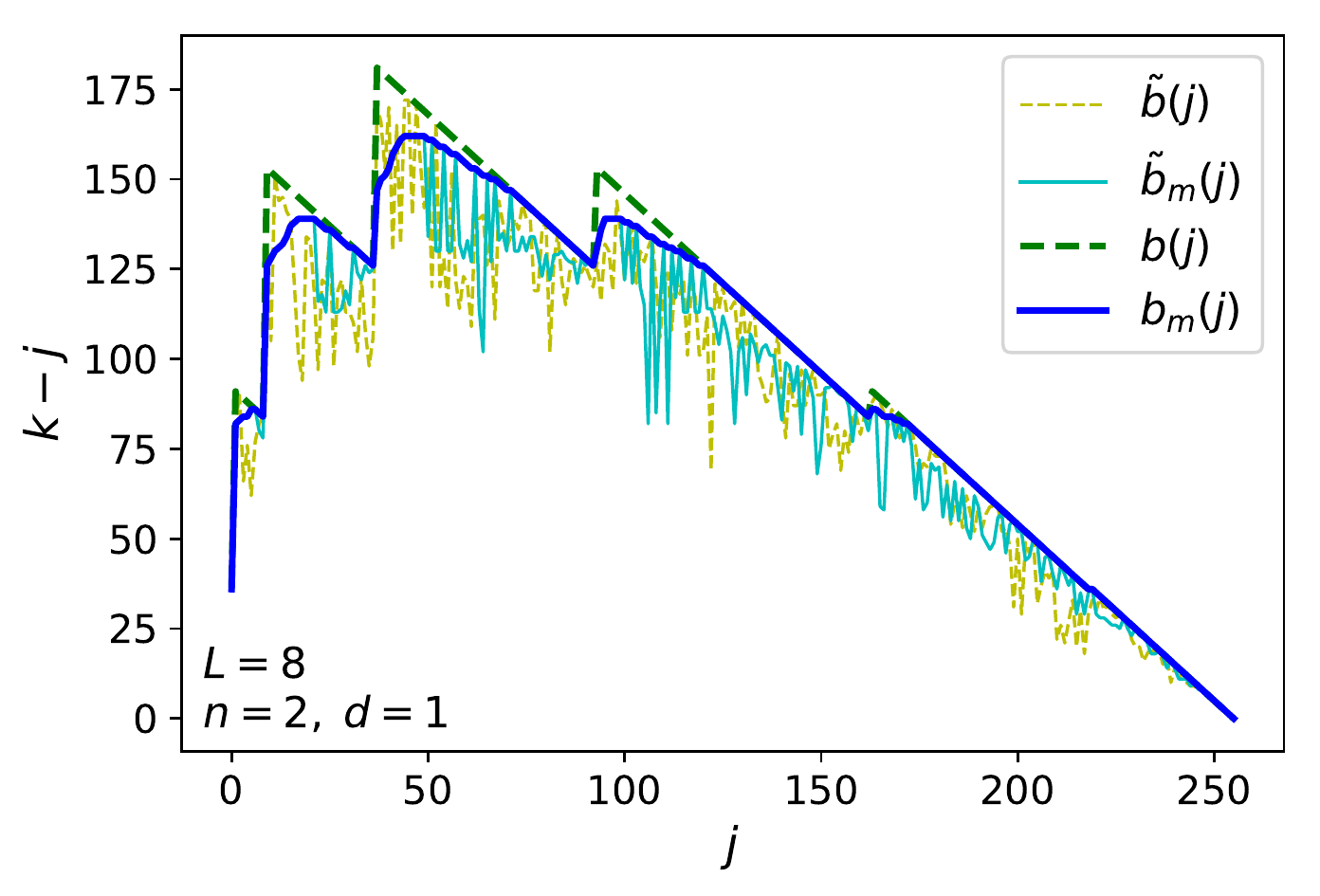}\hfil%
	\includegraphics[width=0.45\linewidth]{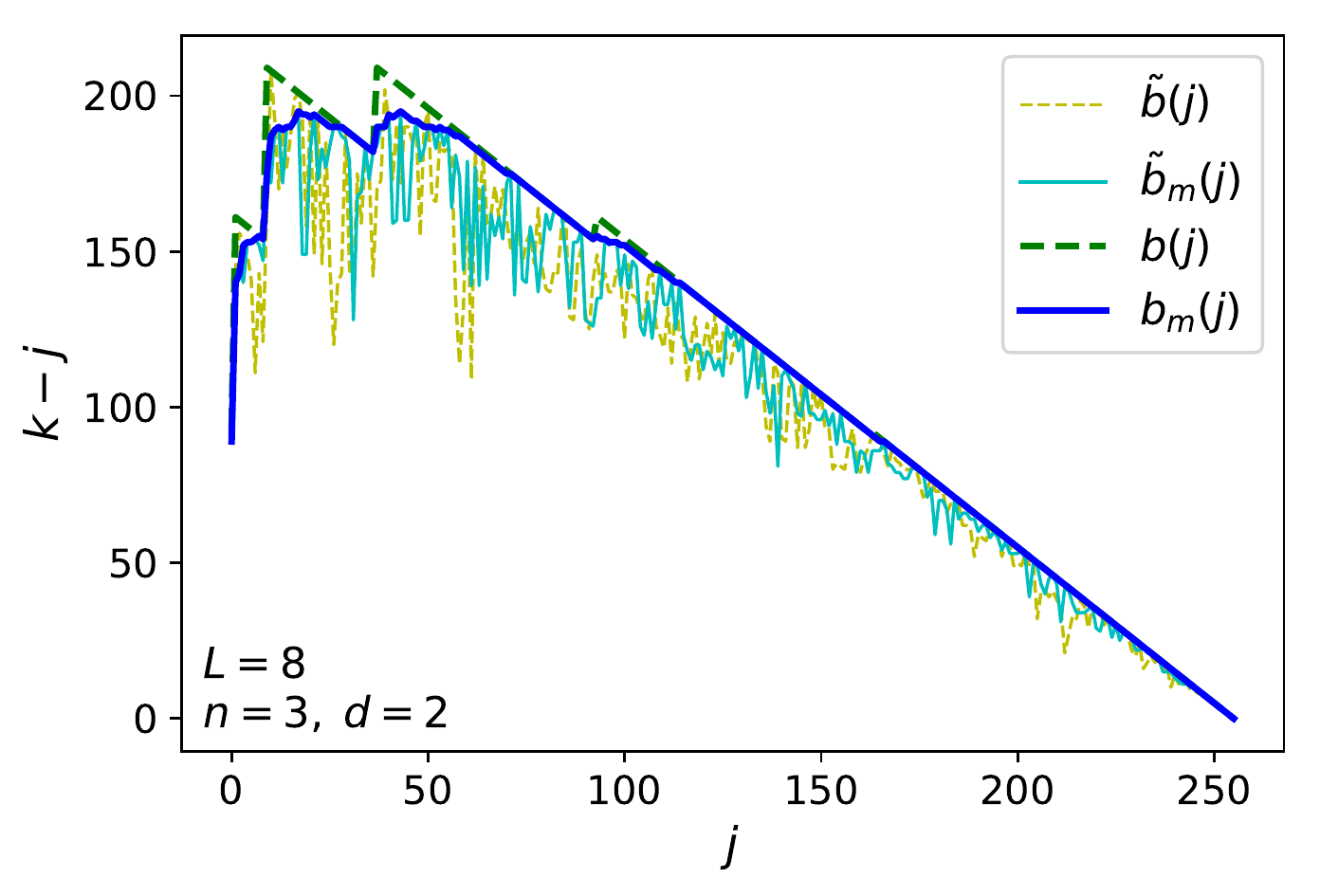}
	\caption{Bandwidth functions $b(j)$ and $b_m(j)$ compared to the actual distance $\tilde b(j)$ and $\tilde b_m(j)$ of the nonzero entry furthest from the diagonal of adjacency matrices $A_{jk}$ from \eqref{eq:adjacency} derived from Hamiltonian in \eqref{eq:H} for $n=2$ and $d=1$ (left) and $n=3$ and $d=2$ (right). The bandwidth $b(j)$ is defined in \eqref{eq:bj} and applies to the case $a_m(i)\equiv1$ for the magnetisation $M$ in \eqref{eq:M} as observable. The bandwidth $b_m(j)$ follows from the general formula \eqref{eq:bo}, here applied to the randomised $M$ using normal-distributed $a_m(i)$ as defined in \eqref{eq:M}. The bandwidths $\tilde b(j)$ and $\tilde b_m(j)$ follow from the definition in \eqref{eq:btilde} for non-randomised and randomised $M$ respectively. \label{fig:bandwidths}}
\end{figure*}

By inspection of the upper rows of Fig.~\ref{fig:masks}, we observe that the bands of nonzero matrix elements are curved, in the sense that in the centre of the matrix nonzero elements occur further away from the diagonal compared to the edges of the matrix. We call this a {\em node-dependent bandwidth}. Since equilibrium states are located towards the centre of the matrix, this implies that a nonequilibrium node $j$ is connected only to nodes with a similar index $i\approx j$, whereas equilibrium nodes may also be connected to nodes whose index $i$ differs more substantially from $j$. For the adjacency matrix of the spin Hamiltonian \eqref{eq:H}, the node-dependent bandwidth can be calculated analytically. We first derive such a result for the magnetisation \eqref{eq:M} with homogeneous weights $a_m(i)\equiv1$, and then generalise the argument to arbitrary observables. 

Writing the Hamiltonian in terms of ladder operators $\s_i^\pm=\frac{1}{2}(\s_i^x\pm i\s_i^y)$, $n$-spin terms translate into products of $n$ ladder operators, which allow for at most $n$ spin flips and can change the magnetisation by at most $\D m=2n/L$. We define the bandwidth
\begin{equation}
b(j)=\max_{k:o_k\leq o_j-\D m}|k-j|
\end{equation}
of node $j$ as the largest distance to any of the nodes $k$ for which $o_k\leq o_j-\D m$. For the magnetisation \eqref{eq:M} with $a_m(i)\equiv1$ there exist $\binom{L}{q}$ states with $q$ up-spins, hence we can write
\begin{align} \label{eq:bj}
b(j) = \sum_{q=0}^n \binom{L}{j+q},
\end{align}
where we assumed for simplicity that node $j$ is the first node of a block of constant magnetisation. Note that $b$ depends on the locality parameter $n$, but not on $d$. Since $H_{jk}$ is typically not densely filled inside the band, the actual distance of the nonzero entry furthest from the diagonal,
\begin{equation}\label{eq:btilde}
\tilde{b}(j)=\max_{k:A_{j-k,j+k}\neq0}|k-j|,
\end{equation}
may be smaller than the bandwidth, $\tilde{b}\leq b$.

Noting that $\binom{L}{q}$ is the density of eigenstates for magnetisation $m=(2q-L)/L$, we can generalize \eqref{eq:bj} to general observables $O$,
\begin{align} \label{eq:bo}
b_o(j) &= \int_{o_j}^{o_j+\D o} g(o) \di o= G(o_j+\D o) - G(o_j).
\end{align}
Here, $g(o)$ is the density of eigenstates $O$, and $G(o)$ the integrated density of states, defined as the number of eigenstates $\kt{o_j}$ of $O$ for which $o_j\leq o$.
By
\begin{align} \label{eq:Do}
\D o = \max\limits_{\psi}\bropkt{\psi}{H^\dagger OH-O}{\psi}
\end{align}
we denote the Hamiltonian's circle of influence in the spectrum of the observable, which can be determined by maximising over the eigenstates $\kt{o_j}$ of $O$. Equation \eqref{eq:bo}, being general, also applies to the magnetisation \eqref{eq:M} with normally distributed weights $a_m(i)$ that we are interested in. Figure~\ref{fig:bandwidths} illustrates how the bandwidths $b(j)$ and $b_m(j)$ compare with the actual structure of the adjacency matrix $A_{jk}$ of the spin model \eqref{eq:H} when using the magnetisation \eqref{eq:M} as observable. Overall, we find that the locality parameter $n$ determines the bandwidth of the adjacency matrix, whereas $n$ and $d$ determine the density of nonzero elements inside the band.

\subsection{Comparison of random graph ensembles}
\label{ssec:ens}
In the previous section we have studied certain graph features of adjacency matrices, in particular their degree distributions $\varrho$ and node-dependent bandwidths $b$, and investigated the dependence of these features on the locality parameters $n$ and $d$ of the Hamiltonian. Now we reverse the logic by constructing random graphs with a certain $b$ and $\varrho$, and investigate to what extend the corresponding random matrix ensembles reproduce the equilibration dynamics of Hamiltonians with locality parameters $n$ and $d$. In the following we sketch the main features of the random graph ensembles considered in this paper. Full descriptions on how the ensembles are constructed are given in Appendices \ref{app:constructionH} and \ref{app:constructionA}.
\begin{description}
	\item{EXH}: {\bf Ex}act {\bf H}amiltonian $H_{jk}$ of the Heisenberg model \eqref{eq:H} with random couplings drawn from a normal distribution with zero mean and standard deviation $1/2$.
	\item{EXA}: Based on the {\bf ex}act {\bf a}djacency matrix $A$ of the Hamiltonian \eqref{eq:H}, random Hamiltonians are created by replacing the nonzero matrix elements of $A$ with normally distributed i.i.d.~numbers.
	\item{BRF}: Randomly generated {\bf b}anded {\bf r}egular graphs with constant degree $\varrho(L,n,d)$ and fixed {\bf f}unctional bandwidths $b_m(j)$. The functional bandwidth \eqref{eq:bo} of the exact Hamiltonian \eqref{eq:H} with a given $n$ and $d$ is used.
	\item{BVF}: Randomly generated {\bf b}anded graphs with {\bf v}ariable degree and fixed {\bf f}unctional bandwidth $b_m(j)$ using \eqref{eq:bo}. The average degree is set to $\varrho(L,n,d)$, and the functional bandwidth matches that of the Hamiltonian \eqref{eq:H} with a given $n$ and $d$.
	\item{BRC}: {\bf B}anded {\bf r}egular graphs with {\bf c}onstant (node-independent) bandwidth $b\equiv \max_j b_m(j)$ using \eqref{eq:bo} are generated randomly but not uniformly. The maximum bandwidth and the constant degree match those of the original Hamiltonian \eqref{eq:H}.
	\item{REG}: {\bf Reg}ular graphs with constant degree $\varrho(L,n,d)$, sampled uniformly. This ensemble is similar to the approach in \cite{Brandino2012}.
\end{description}

\begin{figure}
	\fbox{\includegraphics[width=0.27\linewidth]{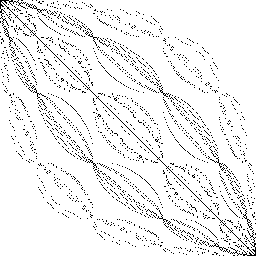}}\quad%
	\fbox{\includegraphics[width=0.27\linewidth]{HEI___Mask_L8_n2_d1.png}}\quad%
	\fbox{\includegraphics[width=0.27\linewidth]{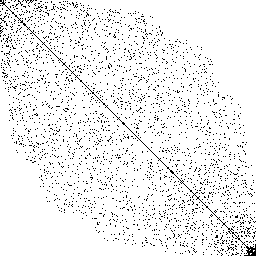}}\\[3pt]
	\hspace{0pt} EXH \hfill EXA \hfill BRF\\[3mm]
	\fbox{\includegraphics[width=0.27\linewidth]{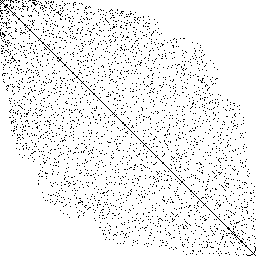}}\quad%
	\fbox{\includegraphics[width=0.27\linewidth]{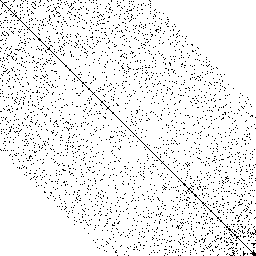}}\quad%
	\fbox{\includegraphics[width=0.27\linewidth]{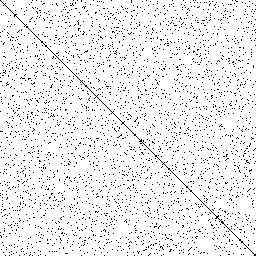}}\\[3pt]
	\hspace{30pt} BVF \hfill BRC \hfill REG
	\caption{Plots of the adjacency matrices for the various random graph ensembles for $L=8$ spins. Note that EXH and EXA are identical since the resulting Hamiltonians differ only in the statistics of their matrix elements, not their adjacencies. \label{fig:graphconstrs}}
\end{figure}

In Fig.~\ref{fig:graphconstrs} we show adjacency matrices for these six random graph ensembles. The graphs of all the ensembles have the same total number of edges, i.e. all resulting Hamiltonians have the same number of nonzero matrix elements. Apart from the EXH ensemble, the weights of all graph edges are i.i.d. numbers sampled from normal distributions matching the statistics of the exact $H_{jk}$. Details of the weight distributions are given in Appendix \ref{app:constructionW}. 

We interpret each realisation of a random graph drawn from one of the above ensembles as the matrix representation $H_{jk}$ of a Hamiltonian in the eigenbasis of the randomised magnetisation $M$ in \eqref{eq:M}. Our aim is to study the equilibration timescales of the Hamiltonians and assess which of the ensembles lead to timescales that faithfully reproduce those of Hamiltonians with the corresponding degrees of localities $n$ and $d$. As discussed in Sec.~\ref{sec:spin}, timescales are made more comparable by normalising each $H_{jk}$ such that $\|H\|=1$. Since equilibrium values of $M$ are expected to be close to zero, we choose the initial state as far away from equilibrium as possible, i.e.\ we select the node $k=N$ that corresponds to a maximum magnetisation.

Using the function \texttt{expm\_multiply} from the python package \texttt{scipy.sparse.linalg}, which is designed for efficient propagation of initial vectors under the linear time evolution of sparse matrices, we compute the exact time evolution for samples of Hamiltonians drawn from each of the above graph ensembles. For each time evolution we measure the equilibration time $T_\text{eq}$, defined as the time at which both $\lla O\rra$ and $\lla O^2\rra$ for the first time reach their diagonal averages with an allowed margin of error of 10\% of their respective initial expectation values \cite{Nickelsen2019}. This definition of $T_\text{eq}$ is in the spirit of the definition  of equilibration in terms of equilibrium subspaces in Sec.~\ref{sec:set}.

\begin{figure*}[t]
	\includegraphics[width=0.32\linewidth]{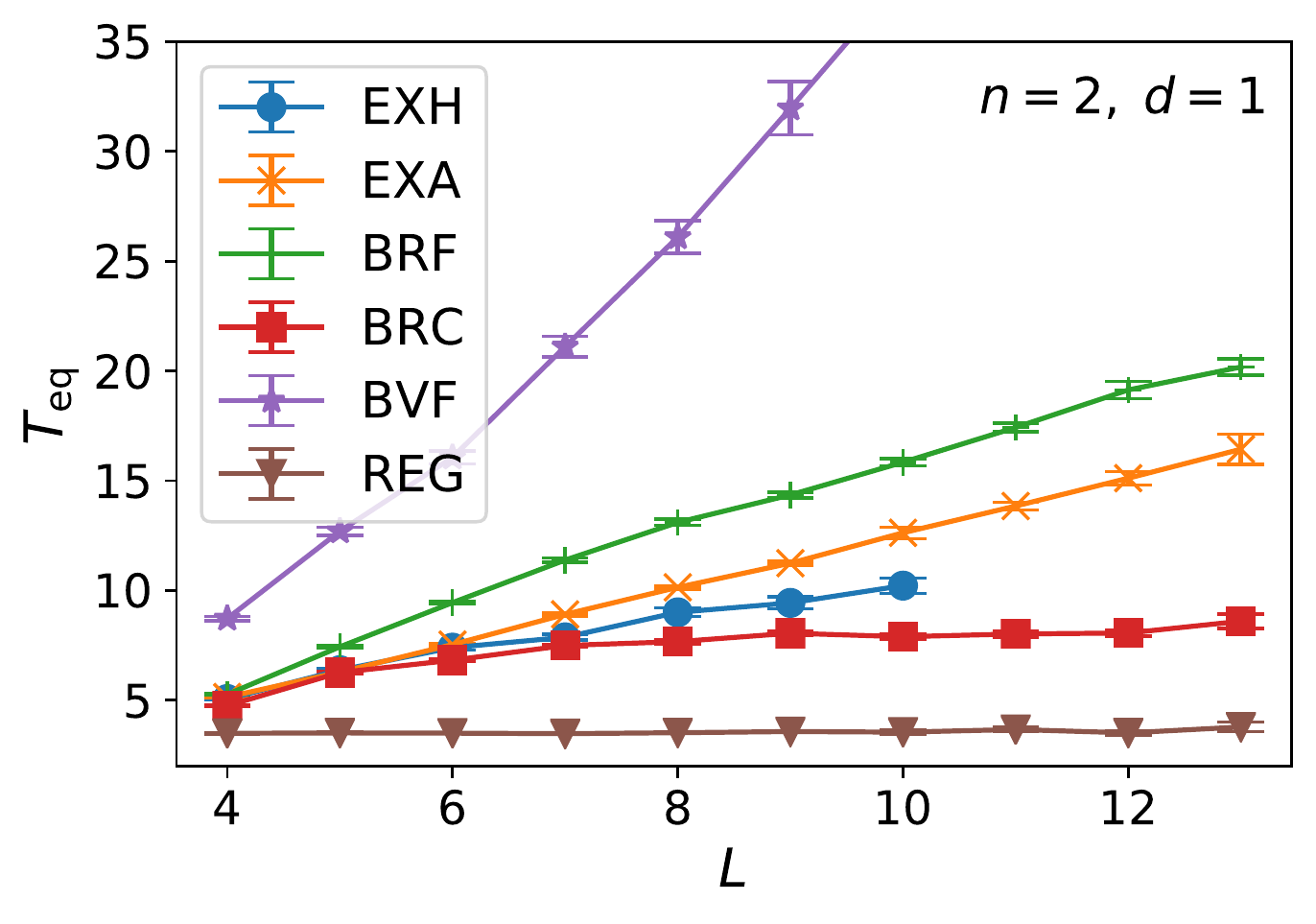}\hfil%
	\includegraphics[width=0.32\linewidth]{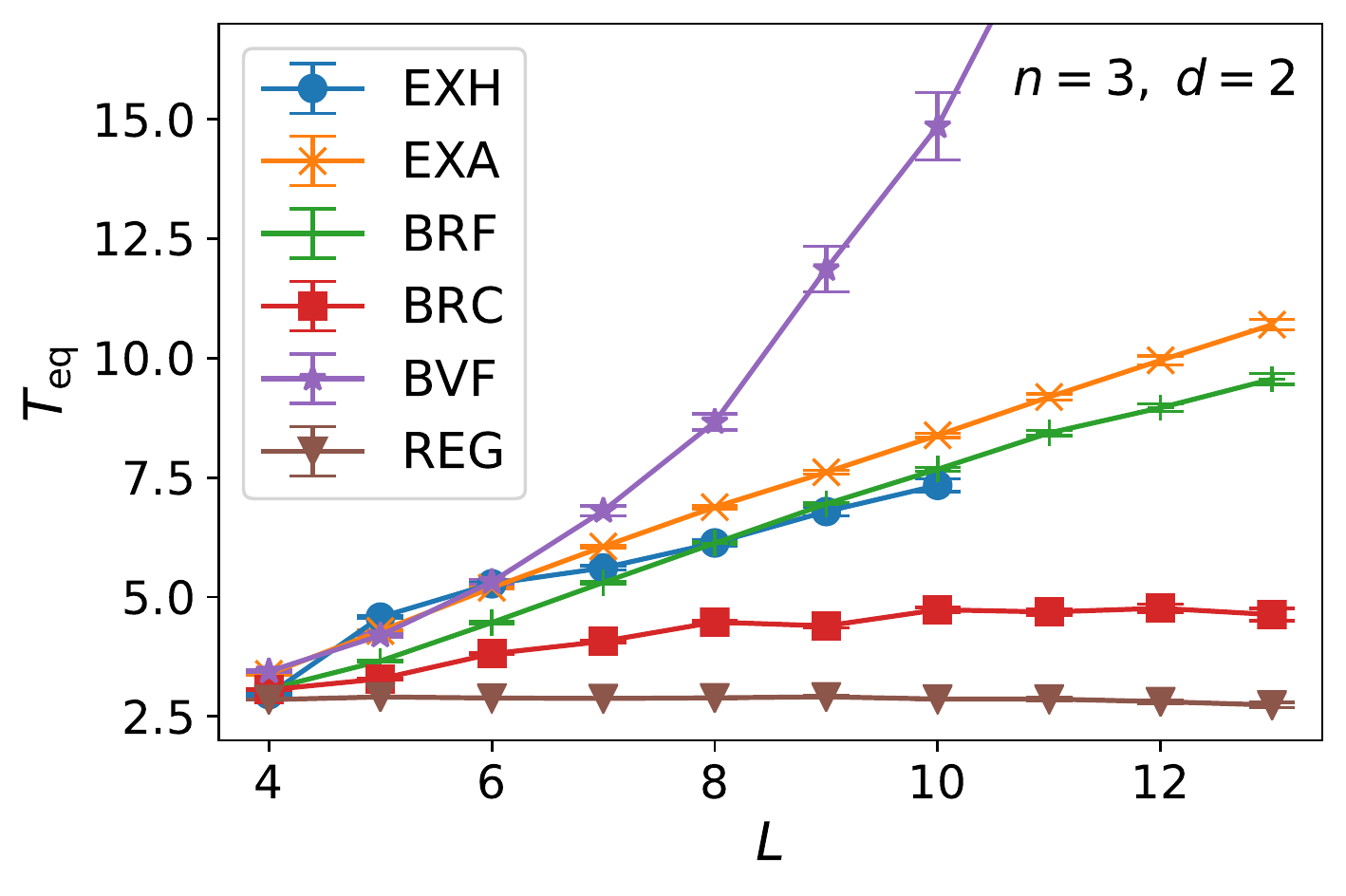}\hfil%
	\includegraphics[width=0.32\linewidth]{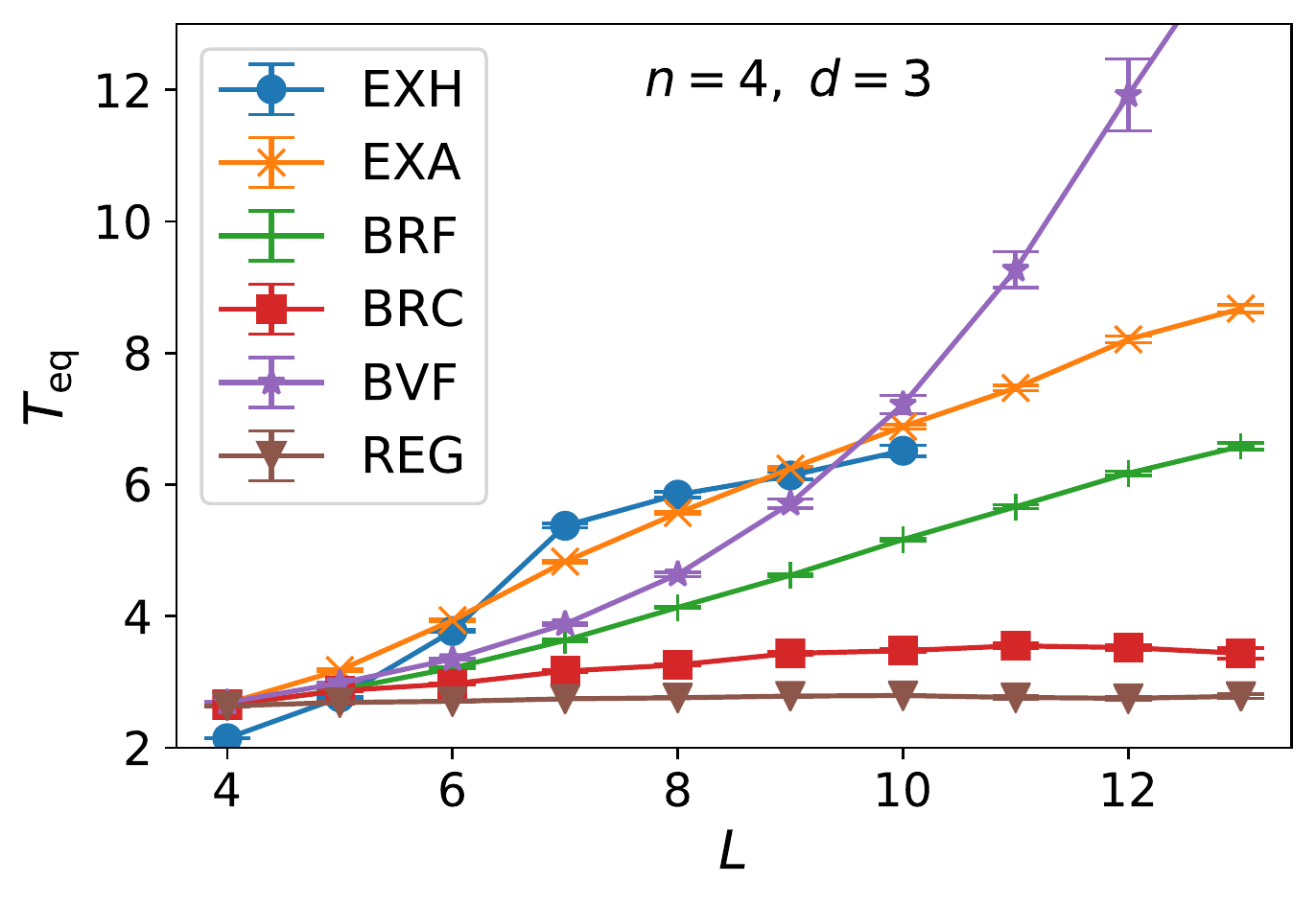}
	\caption{Dependence of the equilibration times $T_\text{eq}$ on the system size $L$ for the six random graph ensembles introduced in Sec.~\ref{ssec:ens}. 
	Averages and standard errors of $T_\text{eq}$ are computed using sample sizes decreasing exponentially as $2^{18-L}$. The smaller sample sizes are justified due to the self-averaging properties of the larger matrices. Different disorder realisations of the randomised magnetisation \eqref{eq:M} are used for all random graphs. Our computational resources restricted us to system sizes of only $L=10$ for the computationally more demanding EXH ensemble. The range of the ordinate focusses on relevant values for $T_\mr{eq}$ for clearer plots. \label{fig:Teqs_masks}}
\end{figure*}

Figure~\ref{fig:Teqs_masks} shows $T_\text{eq}$ for various system sizes. Based on these plots, we make the following observations:
\begin{enumerate}
\renewcommand{\labelenumi}{(\alph{enumi})}
\item The exact ensemble EXH shows an approximately linear increase of $T_\text{eq}$ with the system size $L$ for all three sets of parameters shown in Fig.~\ref{fig:Teqs_masks}.
\item The equilibration times obtained from the REG and BRC ensembles are substantially smaller than those of EXH and do not (or hardly) increase with $L$. This disqualifies REG and BRC as unsuitable for reproducing the equilibration behaviour of local Hamiltonians.
\item The BVF ensemble overestimates the increase of $T_\text{eq}$ with $L$, showing slower equilibration than EXH, at least for the larger system sizes. We speculate that variations in the degrees of nodes cause a bottleneck effect, where poorly connected nodes do not permit efficient propagation of oscillations towards equilibrium nodes.
\item Unsurprisingly, the equilibration times of EXH agree best with those of EXA, the ensemble which has the exact same adjacency matrix. Their Hamiltonian matrix elements, however, differ, even though they are statistically equivalent if taken to be i.i.d.\ numbers. The differences in $T_\text{eq}$ can be attributed to correlations between the matrix elements $H_{jk}$ that have not correctly been accounted for in EXA. This observation is in line with recent studies on the role of correlations between the random matrix elements of the Hamiltonian and the observable \cite{Beugeling2015,Wang2017,Morampudi2018}, a topic that has been discussed also as an extension to the eigenstate thermalisation hypothesis \cite{Nation2018,Foini2019,Brenes2019}. To fully focus on the effects of graph topology, it may make sense to compare equilibration times of the various ensembles to EXA instead of EXH, so that different topologies but identical statistics of matrix elements $H_{jk}$ are compared.
\item Among the ensembles with random adjacency matrices, the BRF ensemble 
is the only ensemble reproducing the linear scaling of $T_\mr{eq}$ with $L$ observed for the ensembles EXH and EXA, and as such best suited to model the equilibration dynamics for the family of Hamiltonians considered in \eqref{eq:H}. However, on the basis of the available data, one may speculate that the quantitative agreement deteriorates with decreasing locality.

\end{enumerate}

We conclude that degree distribution and bandedness are essential features of the adjacency matrices of local Hamiltonians that need to be taken into account in a random matrix approach. The shape of the node-dependent bandwidth also has a role to play, but appears to be somewhat less crucial. Since it is computationally costly to calculate the exact adjacency matrix required for the EXA ensemble, resorting to BRF as the next-best, and computationally less expensive, ensemble is a potential avenue for studying equilibration times of larger systems in a random matrix setting. We do however not claim that the BRF ensemble with its constant degree is always the best choice to model equilibration of local systems. Many other natural Hamiltonians may not be of constant degree (e.g. particular choices of the $a(\chi,\phi)$ in \eqref{eq:H}) -- studying the degree distribution of these systems and how it effects their equilibration timescales should be interesting for future research.

A comment is in order on the increase of $T_\text{eq}$ with system size $L$ that is observed in EXH as well as in several of the approximate random graph ensembles. On physical grounds, one expects a (statistically) homogeneous local system with homogeneous initial conditions to equilibrate on an approximately constant (system-size independent) timescale: because of the homogeneity, no currents across macroscopic distances are required to reach equilibrium, and local equilibration is sufficient to reach global equilibrium. This reasoning applies in the conventional setting of an {\em extensive} local Hamiltonian. Our convention of normalising $H$ such that $\|H\|=1$, however, introduces an extra factor of $1/\|H\|$ in the time-evolution operator, resulting in the observed approximately linear increase of $T_\text{eq}$ with $L$.

\subsection{Maximum flow {\em vs.}\ equilibration time}
\label{ssec:fmax}
To determine the equilibration time for a given Hamiltonian, one needs to integrate the time-dependent Schr\"odinger equation up to sufficiently late times. This is computationally costly and is feasible only for small system sizes. To circumvent this difficulty, we propose to calculate instead the maximum flow value $f_\text{max}$, defined in Sec.~\ref{sec:graph}, of the graph defined by $H_{jk}$, which is numerically less expensive. We show that $T_\text{eq}$ and $f_\text{max}$ are strongly correlated, and by calculating the latter one can estimate the equilibration time to fairly good accuracy. Extrapolating this finding to larger system sizes, estimates of equilibration times of larger systems can be obtained. 

\begin{figure*}[t]
	\includegraphics[width=0.45\linewidth]{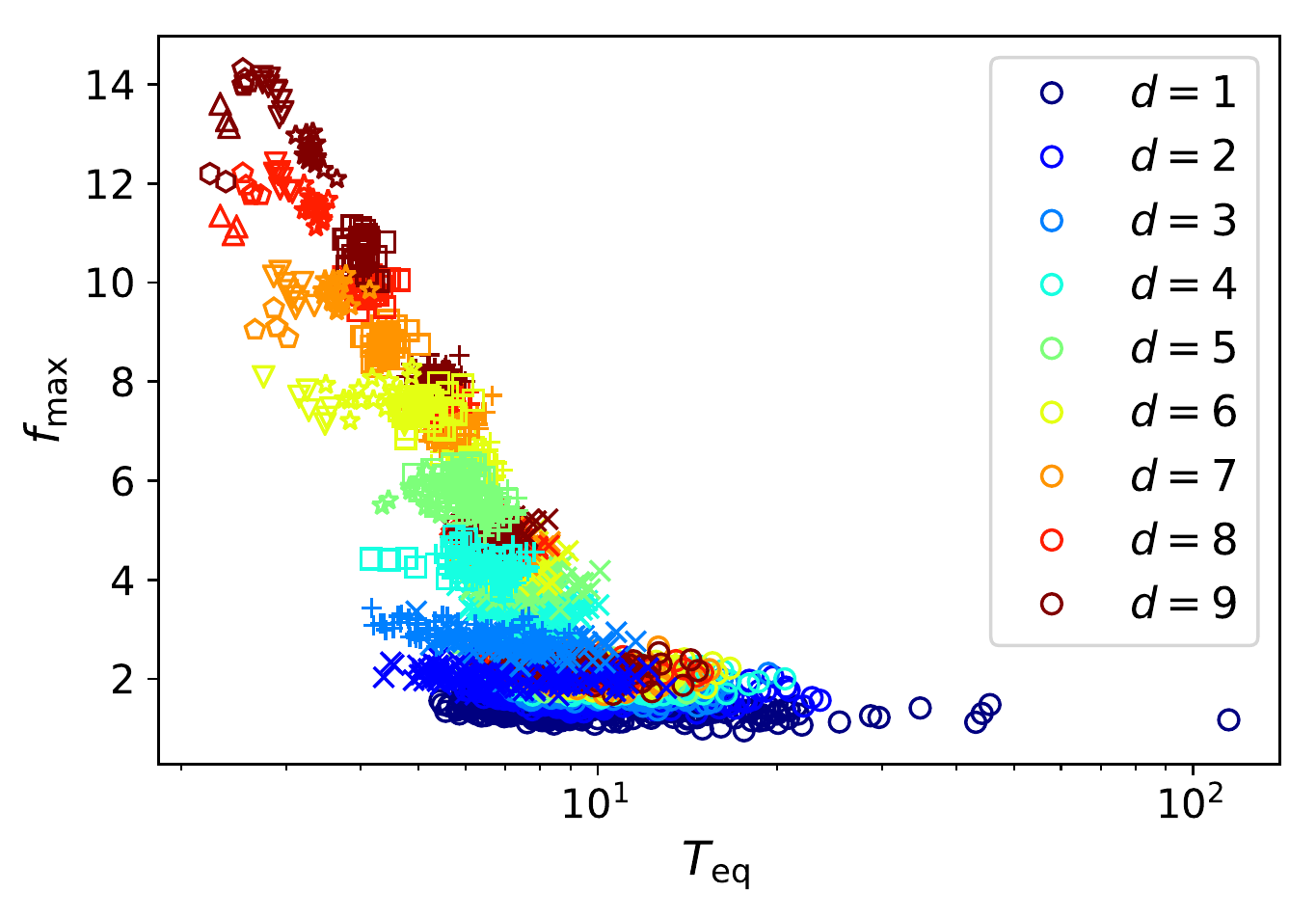}\hfill
	\includegraphics[width=0.45\linewidth]{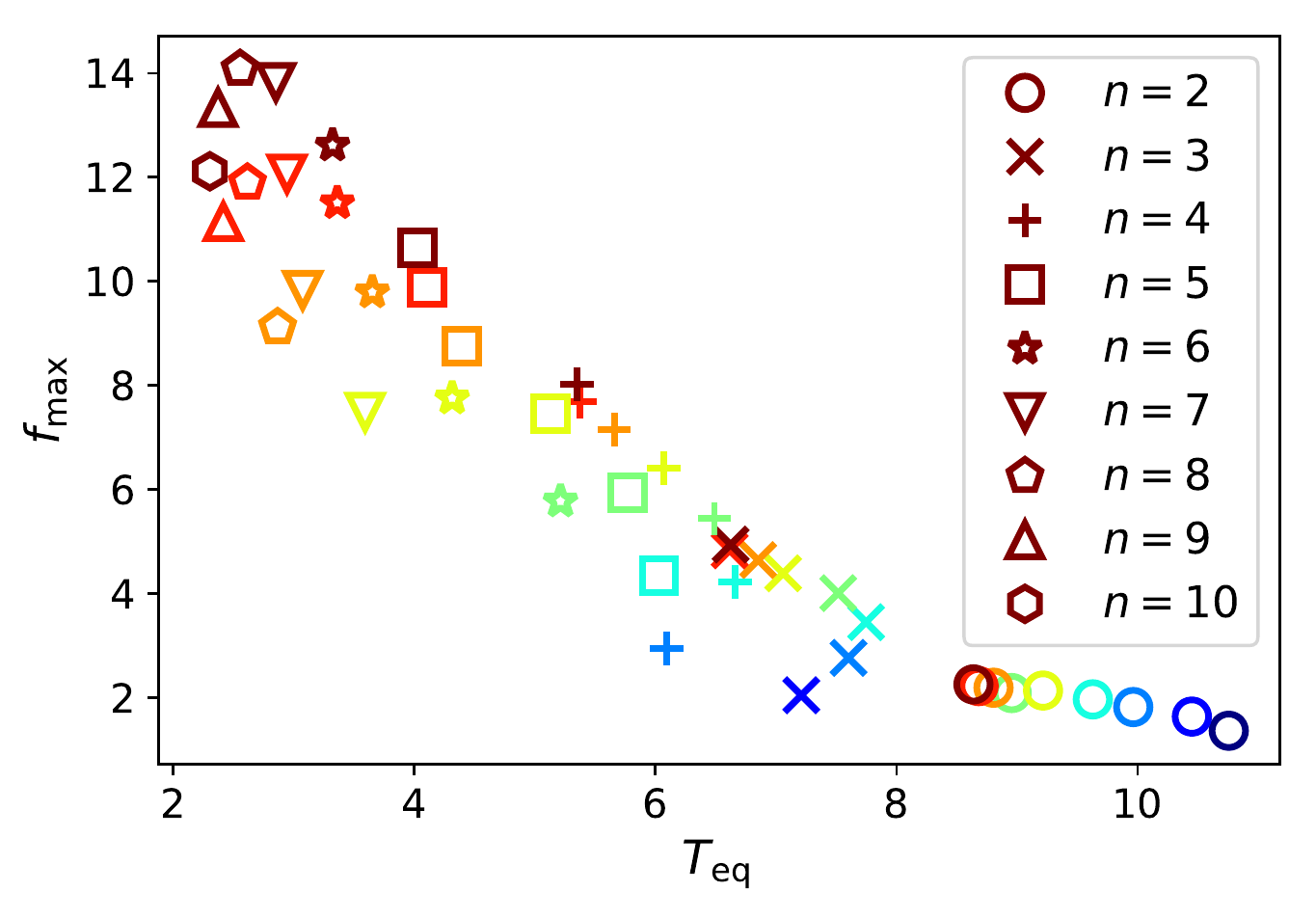}
	\caption{Left: Scatter plot of the maximum flow value $f_\text{max}$ versus equilibration time $T_\text{eq}$ for $L=10$ spins, calculated for all possible combinations of $n=2,\dots,10$ and $d=n-1,\dots,9$. Different symbols correspond to different values for $n$, different colours to different values of $d$ (see both legends). For each combination of $n$ and $d$, $2^{10-n}+1$ instances of $H_{jk}$ from \eqref{eq:H} are used to obtain data sets of pairs $(T_\text{eq},f_\text{max})$. Right: Averages and standard errors of the data in the left plot, separately for each combination of $n$ and $d$ (standard errors are smaller than the symbol sizes). \label{fig:Teqs_flowmax}}
\end{figure*}

We use the \texttt{preflow\_push} algorithm of the python package \texttt{NetworkX} to calculate $f_\text{max}(k,j)$ for various realisations of the spin Hamiltonian \eqref{eq:H}. In Fig.~\ref{fig:Teqs_flowmax} (left) we show a scatter plot of the pairs $(f_\text{max},T_\text{eq})$ for various values of $n$ and $d$, and in Fig.~\ref{fig:Teqs_flowmax} (right) the corresponding averages and standard deviations. The plots show strong anticorrelations between $f_\text{max}$ and $T_\text{eq}$: shorter equilibration times are observed when the maximum flow value is larger, and {\em vice versa}. Small imperfections in the anticorrelations may point towards limitations of the usage of $f_\text{max}$ as a proxy for $T_\text{eq}$, or they may simply be caused by shortcomings of our numerical method for computing equilibration times.

\begin{figure*}
	\includegraphics[width=0.45\linewidth]{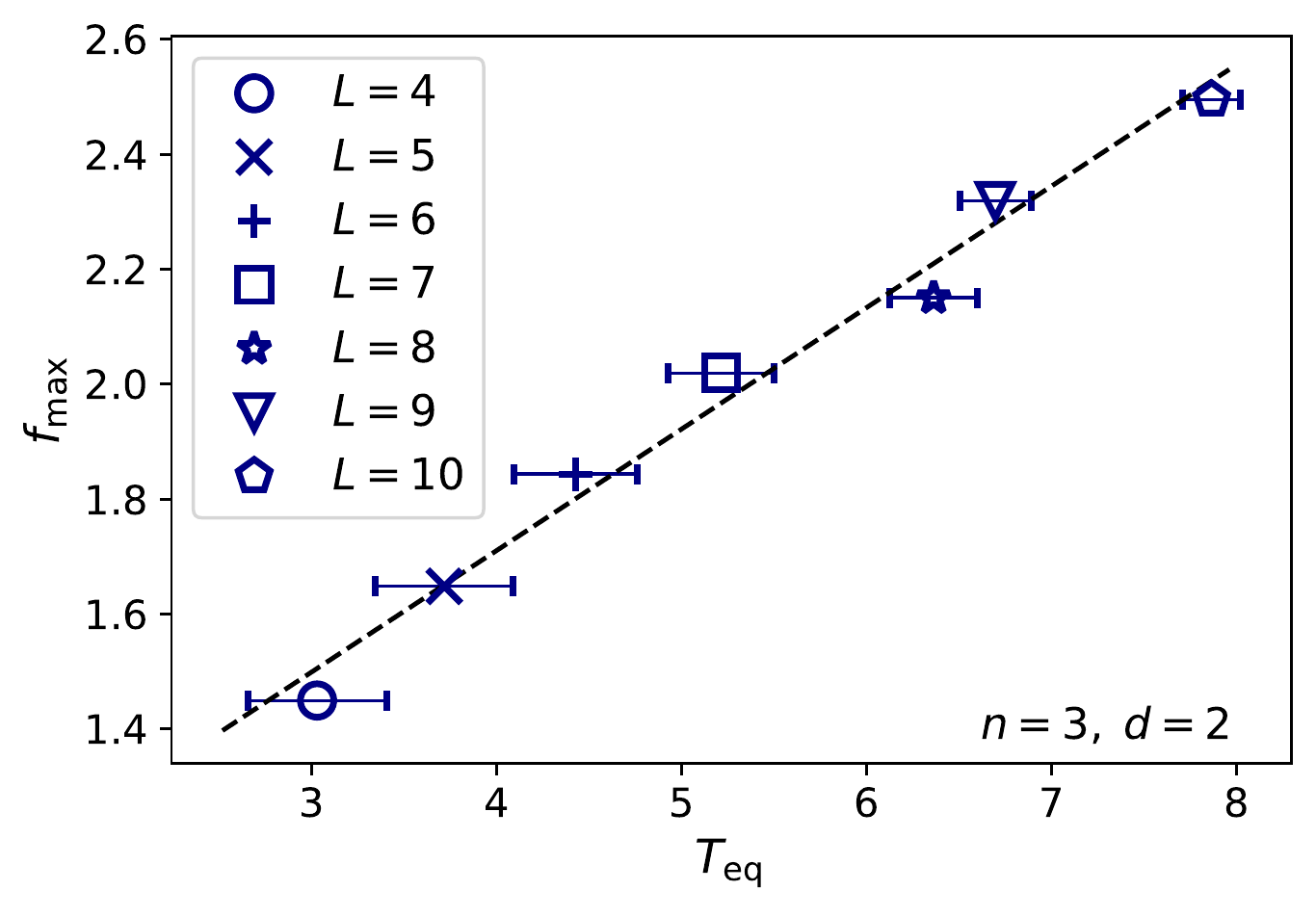}\hfill
	\includegraphics[width=0.45\linewidth]{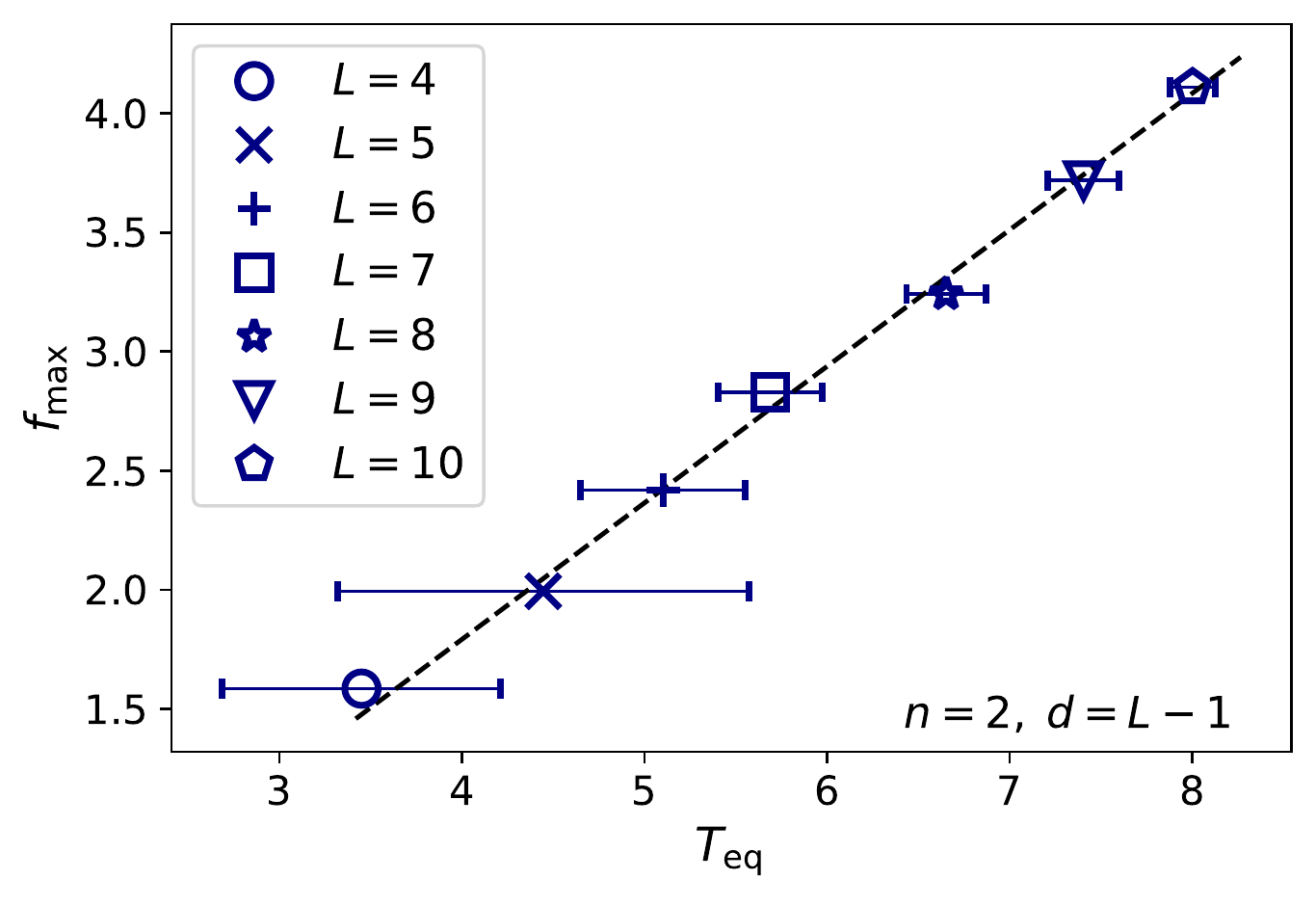}%
	\caption{Sample averages of maximum flow values $f_\text{max}$ versus equilibration times $T_\text{eq}$ for random Hamiltonians from the BRF ensemble. 
	Averages and standard errors of $2^{14-L}$ samples are shown, where the standard errors for $f_\text{max}$ are smaller than the symbol sizes. The straight lines are linear fits to the data. \label{fig:Teqs_flowmax_L}}
\end{figure*}

$f_\text{max}$ and $T_\text{eq}$ scale differently with system size $L$. To use $f_\text{max}$ as an estimate for $T_\text{eq}$ for larger systems, we fix $n$ and $d$ and calculate $f_\text{max}$ and $T_\text{eq}$ for random samples of Hamiltonians for system sizes between 4 and 10. In Fig.~\ref{fig:Teqs_flowmax_L} we show results for random Hamiltonians from the BRF ensembles corresponding to $(n=3,d=2)$ and $(n=2,d=L-1)$. The data suggest a linear relation between $T_\text{eq}$ and $f_\text{max}$. A linear fit to the data, shown in the plot, is used in Sec.~\ref{ssec:esti} to estimate equilibration times of larger systems.

\subsection{Equilibration times of larger systems}
\label{ssec:esti}
\begin{figure}[t]
	\includegraphics[width=0.9\linewidth]{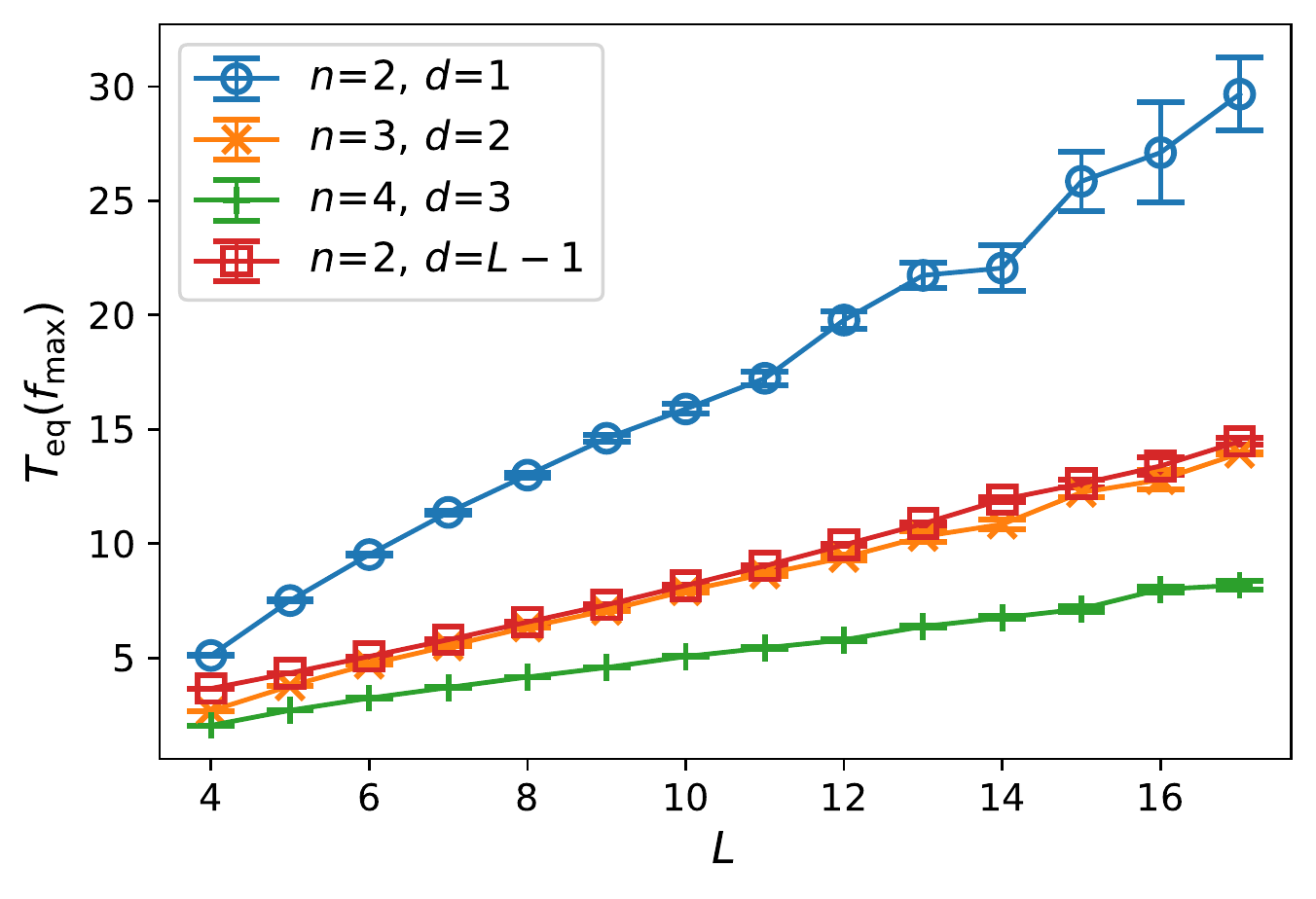}%
	\caption{Equilibration times $T_\text{eq}$ obtained from maximum flow values $f_\text{max}$ of random graphs of the BRF ensemble for various system sizes $L$. Different disorder realisations of the randomised magnetisation \eqref{eq:M} are used for all random graphs. The relation between $T_\text{eq}$ and $f_\text{max}$ is given by linear fits as shown in Fig.~\ref{fig:Teqs_flowmax_L}. Displayed are averages and standard errors of $\max(4,2^{18-L})$ samples for each $L$. \label{fig:flowmax_goL}}
\end{figure}

Based on the results of Secs.~\ref{ssec:ens} and \ref{ssec:fmax}, we are now in the position to sample Hamiltonians from the BRF ensemble for different locality parameters, compute the maximum flow values of the corresponding graphs, and use the linear extrapolations of Fig.~\ref{fig:Teqs_flowmax_L} to deduce $T_\text{eq}$ for system sizes up to $L=17$. While these are still small systems, they are larger than those that can be treated directly with the computational resources at our disposal. In Fig.~\ref{fig:flowmax_goL} we plot the thus obtained results for system sizes between 4 and 17, and for various combinations of locality parameters $n$ and $d$. As expected, $T_\text{eq}$ increases with $L$, and larger values of $n$ entail faster equilibration.

The limitation to $L\leqslant17$ is due to the calculation of $f_\text{max}$, which has a complexity of $O(\sqrt{\varrho}N^{5/2})$ when using the highest-label preflow-push algorithm implemented in \texttt{NetworkX}. The algorithm used to generate the random graphs in the BRF ensemble (see Appendix \ref{app:constructionA}) has an estimated complexity of $O(\varrho N)$, with potential for improvement.

\section{Conclusions}
\label{sec:concl}
We introduced and studied various random graph ensembles, with the aim of modelling equilibration in a spin system with a given degree of locality. Interpreting the Hamiltonian matrix elements as the weights of a graph in Hilbert space, we employed concepts and tools from classical network theory to study quantum mechanical time evolution. The rationale of our analysis proceeds in three steps:
\begin{enumerate}
\renewcommand{\labelenumi}{(\alph{enumi})}
\item Working with a local Hamilton $H$ and a local observable $O$ implies the existence of a basis in which $H$ and $O$ are simultaneously sparse. A numerical study of quantum spin models with random couplings revealed further manifestations of locality in the matrix structure, in particular a locality-dependent constant degree \eqref{eq:degdist} and a node-dependent bandwidth \eqref{eq:bj}.
\item Based on these observations we introduce the random graph ensembles EXA, BRF, BRC, BVF, and REG, which incorporate degree distribution and bandwidth to a certain extent. We numerically study samples of Hamiltonians from the various ensembles, finding that the BRF ensemble faithfully captures the dependence of equilibration times on locality, while having computational advantages over the ensemble EXH of exact random Hamiltonians.
\item To avoid the costly numerical computation of equilibration times, we showed that the maximum flow value $f_\text{max}$ between nonequilibrium and equilibrium nodes of a graph is strongly correlated with the equilibration time $T_\text{eq}$ of the corresponding Hamiltonian. Calculating $f_\text{max}$ and inferring $T_\text{eq}$ is computationally less costly and gives access to equilibration times for system sizes that cannot be reached by direct numerical integration.
\end{enumerate}

Even though we started out from the assumption that the random graphs represent the Hamiltonian in the eigenbasis of the observable, the construction of the random graph ensemble is not specific to a single observable: only the observable's density of states $g(o)$ and the Hamiltonian's circle of influence $\D o$ \eqref{eq:Do} enter the construction, and all observables that share these features can be modelled by the same random graph ensemble. The effect of $g(o)$ and $\D o$ on equilibration times for large system sizes can then be studied using $f_\text{max}$ of the generated graphs, without deriving exact Hamiltonians, spectral properties, or solving the Schr\"odinger equation.

Presently, the modelling of local systems by random graph ensembles, and also the estimation of equilibration times through maximum flow values, led to only a moderate increase of the system sizes we can deal with, but there is plenty of potential for further improvement. The method for constructing banded random adjacency matrices with constant degree described in Appendix \ref{app:constructionA} is certainly far from optimal, and also does not sample uniformly, which may affect the accuracy of the modelling. Also, graph measures other than the maximum flow value may be more suitable and are worth being investigated. We have looked into a few others, including the node connectivity and the shortest path length, but these turned out to be less strongly correlated with $T_\text{eq}$.

Finally, we believe that the random graph ensembles introduced in this paper, as well as the network-theoretic tools employed to analyse the random graphs, have potential for the analysis of local quantum systems well beyond the specific question of equilibration addressed here. It should be interesting to study the effects of locality on the level spacing statistics, a topic that we are planning to report on in future work.

\begin{acknowledgements}
The authors gratefully acknowledge helpful comments on the manuscript from Fausto Borgonovi. M.\,K.\ acknowledges financial support from the National Research Foundation of South Africa through the Incentive Funding and the Competitive Programme for Rated Researchers.
\end{acknowledgements}

\appendix

\section{More on the network perspective}
\label{app:netequi}
In Ref.~\cite{Nickelsen2019} it has been argued that the quantity
\begin{equation}\label{eq:LR_obj}
\begin{split}
\Lambda_{jk}(t) &= \bigg|\frac{\pt x_j(t)}{\pt x_k(0)}\bigg| = \Big|\big(e^{-iHt}\big)_{jk}\Big| \\
&= \bigg|\sum_{q=0}^\infty \frac{\big(-i\|H\|t\big)^q}{q!}\, \frac{(H^q)_{jk}}{\|H\|^q} \bigg|,
\end{split}
\end{equation}
can be used to measure the speed of equilibration. $\Lambda_{jk}(t)$ quantifies the influence a nonequilibrium node $k$ at time zero has on an equilibrium node $j$ at time $t$. Typically, $\Lambda_{jk}(t)$ is small for small values of $t$, and increases with time until it saturates. A large value of $\Lambda_{jk}$ indicates that a change in $x_k$ affects $x_j$ after a time $t$. If the graph distance between nodes $k$ and $j$ of the network is maximal, then \emph{any} initial state can excite the equilibrium node $x_j$, as all other nodes are closer to $j$. Therefore, when $\Lambda_{jk}(t)$ saturates close to its maximal values, we can say the system has equilibrated.

\begin{figure*}[t]
	\includegraphics[width=0.45\linewidth]{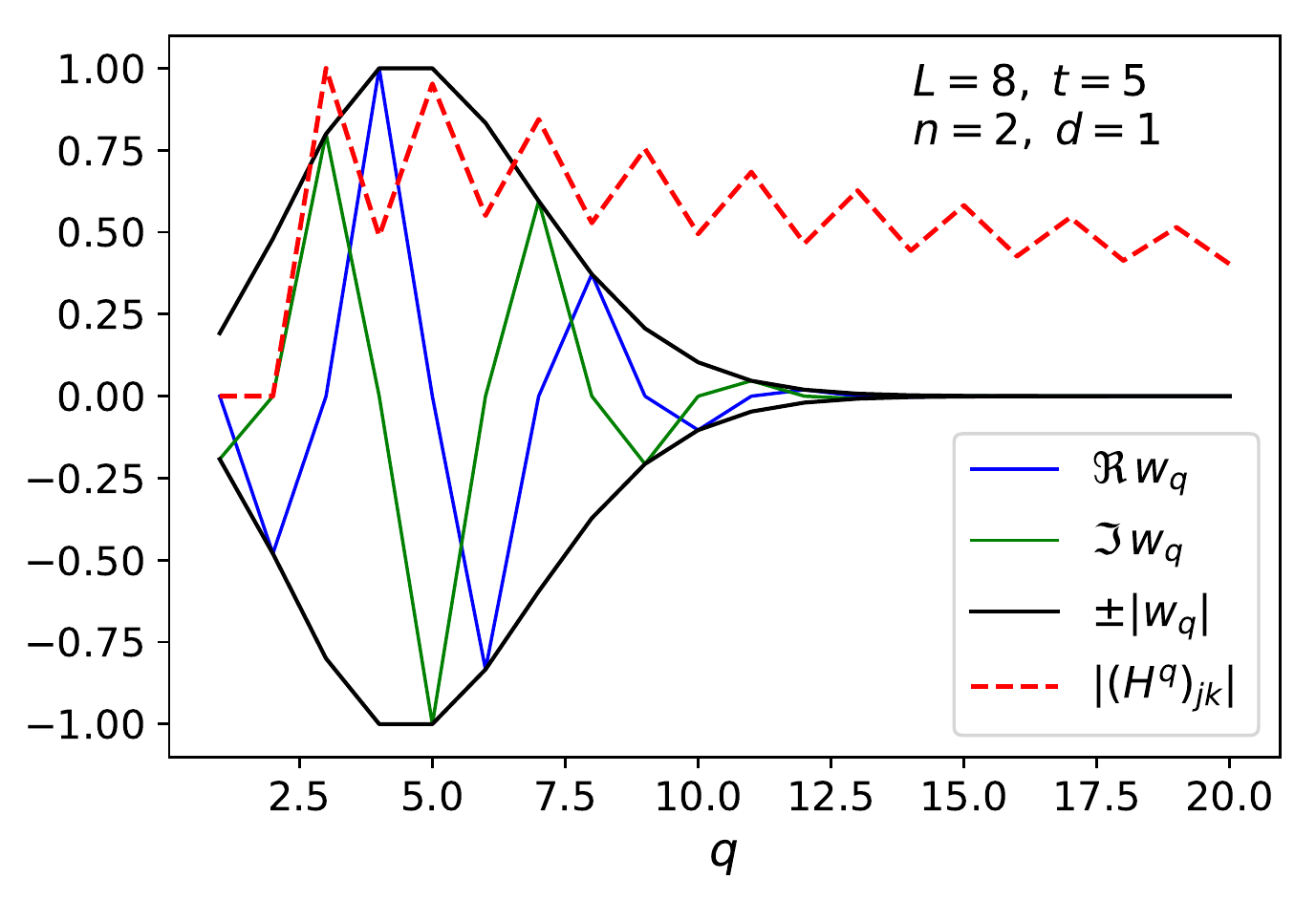}\hfill	\includegraphics[width=0.45\linewidth]{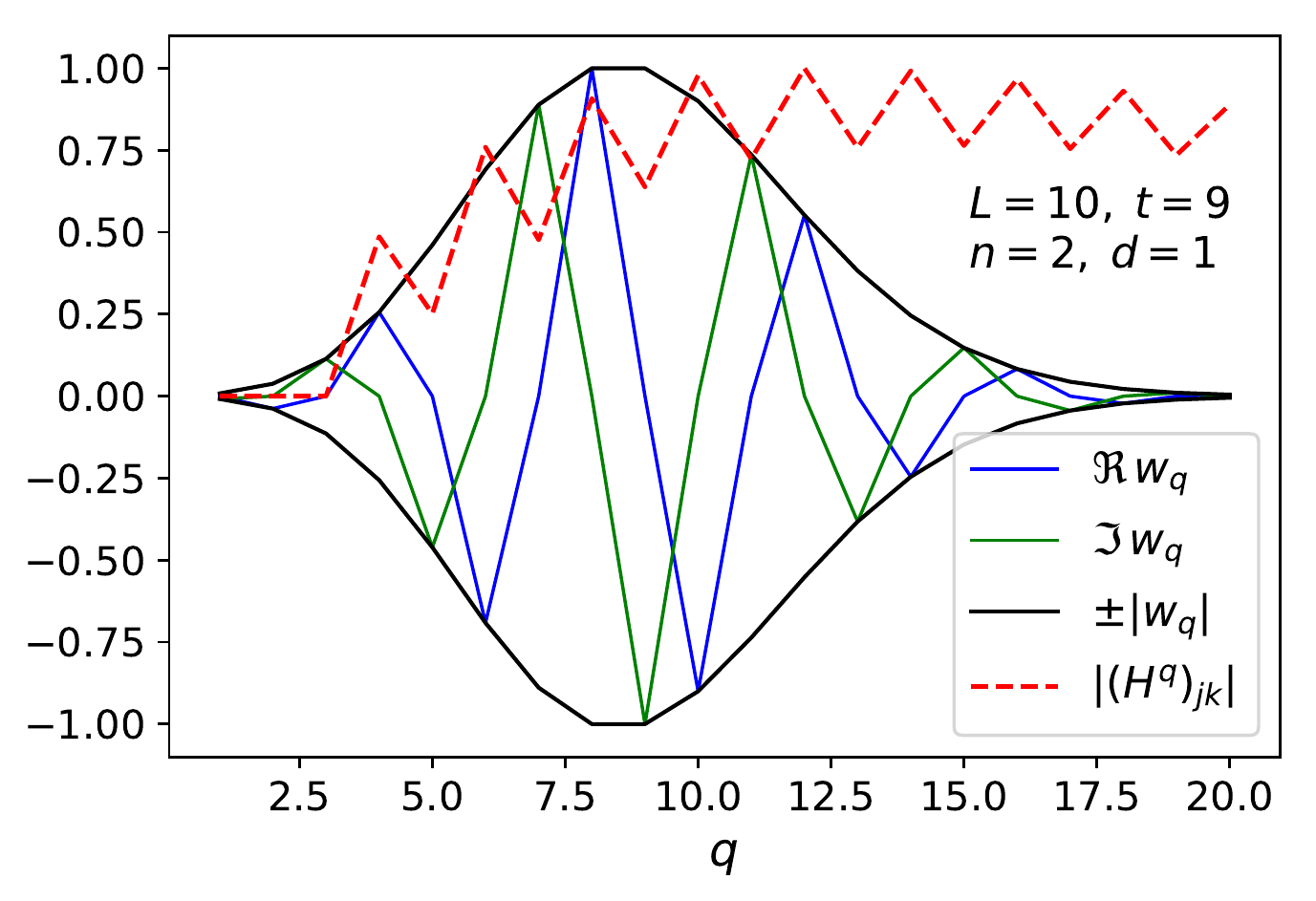}\\
	\includegraphics[width=0.45\linewidth]{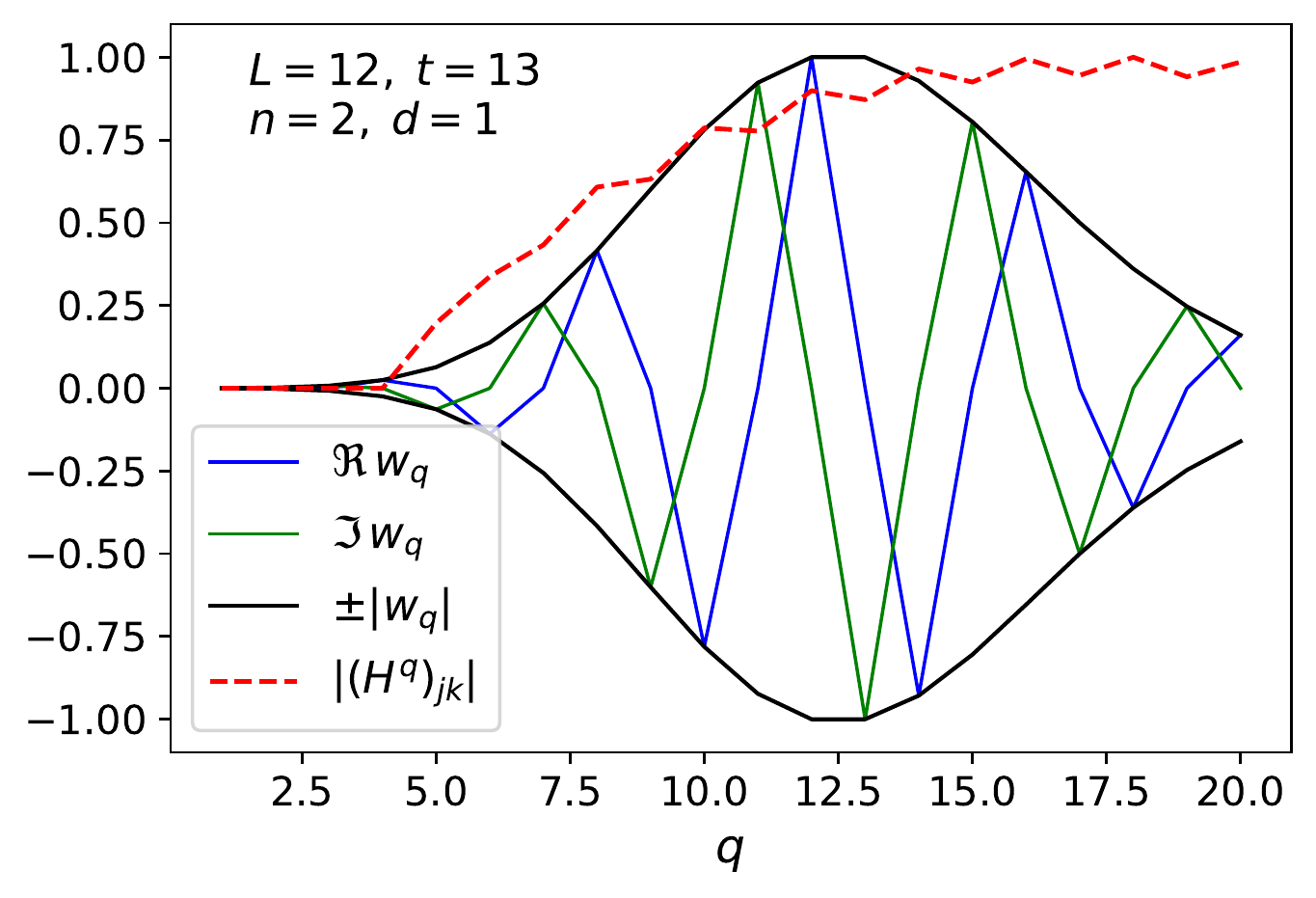}\hfill	\includegraphics[width=0.45\linewidth]{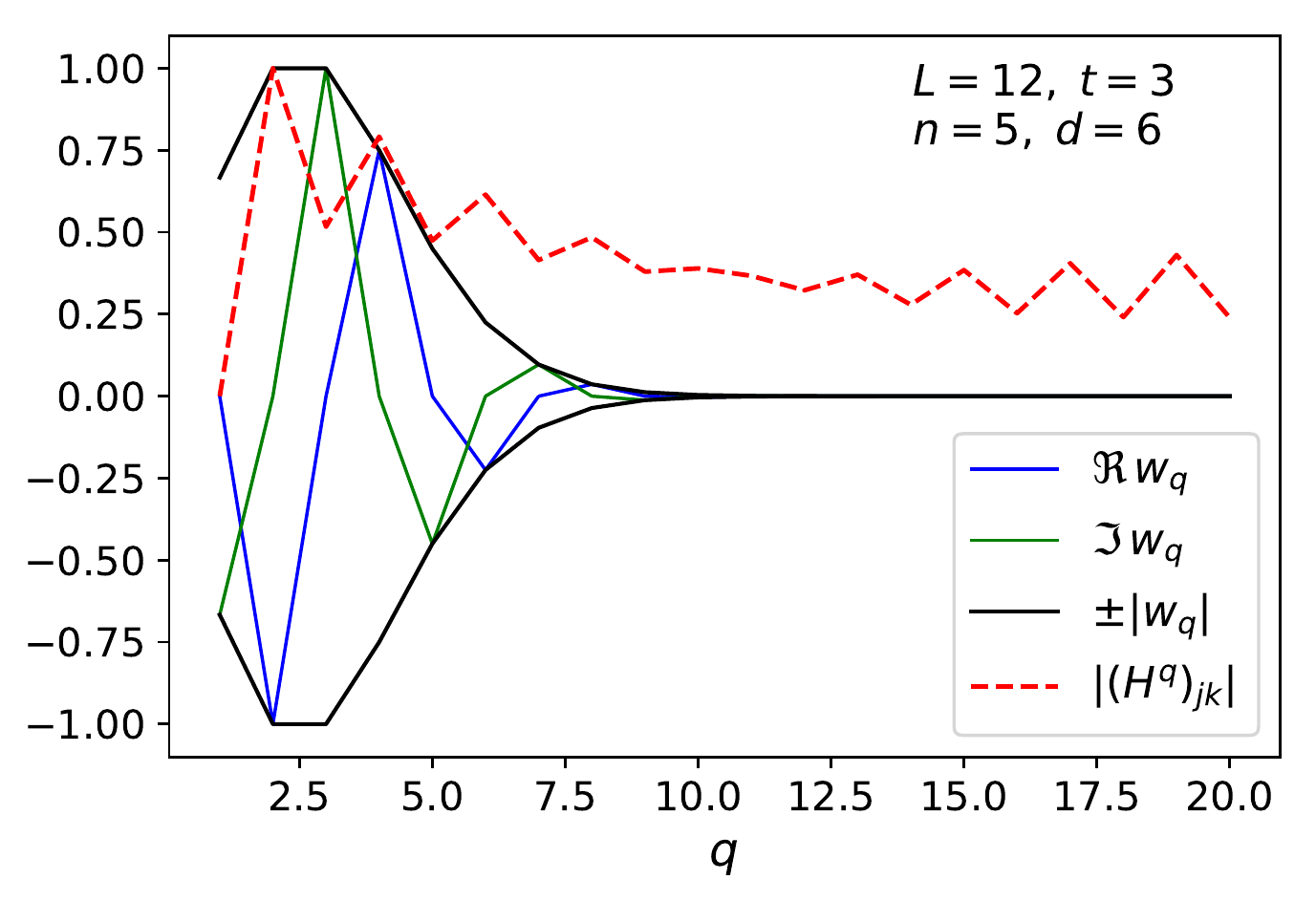}
	\caption{\label{fig:graphint}
	Illustration of the terms in \eqref{eq:LR_obj} with $w_q=(-it)^q/q!$, and $H$
from the EXA ensemble introduced in Sec.~\ref{ssec:ens} and normalised such that $\|H\|=1$. As indicated in the plots, the data are for different numbers of spins $L$, times $t$, and locality parameters $n$ and $d$ as defined in Sec.~\ref{sec:spin}. The $(H^q)_{jk}$ and the $|w_q|$ are divided by their respective maxima for clarity.
}
\end{figure*}

It is insightful to have a closer look at the expansion \eqref{eq:LR_obj} in powers of $H$. The matrix components of these powers are of the form
\begin{align}
(H^q)_{jk} &= \sum\limits_{i_1,\ldots,i_{q-1}} H_{ji_1} H_{i_1i_2} \cdots H_{i_{q-2}i_{q-1}} H_{i_{q-1}i_k} \nonumber\\
&= \sum_{\omega_{jk}(q)} \prod_{\kp\in \omega_{jk}(q)} H_\kp = \sum_{\omega_{jk}(q)} \mathscr{C}(\omega_{jk}(q)), \label{eq:Hn}
\end{align}
where the sums in the second line run over all walks $\omega_{jk}(q)$ that connect nodes $j$ and $k$ using $q$ steps, e.g. $\omega_{25}(4)\in\{\{(24),(43),(34),(45)\},\,\ldots\}$. Walks are allowed to visit the same nodes multiple times. To each walk we assign the corresponding {\em coupling chain}
\begin{equation}
\mathscr{C}(\omega_{jk}(q))=\prod_{\kp\in \omega_{jk}(q)} H_\kp,
\end{equation}
and hence Eq.~\eqref{eq:Hn} expresses the $q$th power of $H$ as the sum over all possible coupling chains of length $q$. For sparse matrices $H$, all small-$q$ coupling chains between far-apart nodes may be zero. We can then interpret the expansion \eqref{eq:LR_obj} as the weighted sum of all coupling chains $\mathscr{C}(\omega_{jk}(q))$ between nodes $j$ and $k$, with $q$-dependent weights $w_q=(-i\|H\|t)^q/q!$. For $t=0$, nodes $j$ and $k$ are uncoupled. For small times $t<1/\|H\|$, $|w_q|$ peaks for couplings of short length ($q\approx2$), and $(H^q)_{jk}$ contributes to the sum only for small values of $q$. Once $t>1/\|H\|$, peak values of $|w_q|$ shift to longer coupling chains ($q\gg1$), and $(H^q)_{jk}$ with larger $q$ contribute to the sum. Since the statistics of the $(H^q)_{jk}$ saturates for sufficiently large $q$, and due to the alternating signs of $(-i)^q$, the value of $\Lambda_{jk}(t)$ oscillates around a stable average for sufficiently large $t$. This is the saturation mentioned above, which indicates that the system has equilibrated. 

The inverse of the norm of the Hamiltonian, $1/\|H\|$, therefore gives an indication on the equilibration timescale, but it does not capture the influence of locality. Locality enters through the vanishing coupling chains of small lengths $q$. For the dense Hamiltonian of a nonlocal system, the weights $w_q$ on short coupling lengths $q\approx2$ for small $t$ are already sufficient for the fluctuations of $\Lambda_{jk}(t)$ to saturate. For the sparse Hamiltonian of a local system, however, coupling lengths are longer and increase with system size, and $t$ needs to be sufficiently large in order for the $w_q$ to give a nonnegligible weight to these coupling lengths, resulting in longer equilibration times. 

In Fig.~\ref{fig:graphint} we illustrate how the two factors $w_q$ and $(H^q)_{jk}$ that enter the expansion \eqref{eq:LR_obj} depend on $q$ for the Hamiltonian \eqref{eq:H} with $n$-spin interactions over a maximum range of $d$ lattice sites. For nearest-neighbour pair-interactions ($n=2$, $d=1$) and $L=8$ spins, coupling chains of length $q\geq2$ contribute. For a larger system of $L=12$ spins only coupling chains of length $q\geq4$ contribute, for which $w_q$ is negligible for $t\approx\|H\|$. For less local choices of parameters ($n=5$, $d=6$), $|w_q|$ already peaks at $q=2$, bringing the system close to equilibrium for $t\approx\|H\|$.

\section{Constant degree}
\label{app:degree}
As a consequence of statistical isotropy and homogeneity of the Hamiltonian, all nodes of a given adjacency graph have a constant degree $\varrho$. By numerically determining $\varrho$ for system sizes up to $L=16$ and various combinations of locality parameters $n$ and $d$, we conjectured that
\begin{align} \label{eq:degfct}
\varrho(L,n,d) = L\sum_{q=0}^{n-1} \binom{d}{q} - C(n,d),
\end{align}
where the first term depends linearly on $L$, and the not yet determined second term $C(n,d)$ is independent of $L$. By studying $C$ for various fixed values of $n$ we were able to further conjecture that
\begin{subequations}
\begin{align}
C(2,d) &= \sum_{i=0}^{d-1}\bigg(\sum_{j=1}^i 1\bigg) , \\
C(3,d) &= \sum_{i=0}^{d-1}\bigg(\sum_{j=0}^i \bigg(1+\sum_{k=1}^j 2\bigg)\bigg) , \\
C(4,d) &= \sum_{i=0}^{d-1}\bigg(\sum_{j=0}^i \bigg(1+\sum_{k=1}^j \bigg(2+\sum_{l=2}^k 3\bigg)\bigg)\bigg),\label{eq:C4} 
\end{align}
\end{subequations}
and so on. Extrapolating the observed pattern to arbitrary $n$ we obtain
\begin{widetext}
\begin{align}
C(n,d) &= \sum_{j_1=0}^{d-1}\sum_{j_2=0}^{j_1}1 + 2\sum_{j_1=0}^{d-1}\sum_{j_2=0}^{j_1}\sum_{j_3=1}^{j_2}1 + 3\sum_{j_1=0}^{d-1}\sum_{j_2=0}^{j_1}\sum_{j_3=1}^{j_2}\sum_{j_4=2}^{j_3}1 +\dots+ (n-1)\sum_{j_1=0}^{d-1}\sum_{j_2=0}^{j_1}\sum_{j_3=1}^{j_2}\dots\sum_{j_n=n-2}^{j_{n-1}}1\nonumber\\
&= \sum_{j_1=0}^{d-1}\sum_{j_2=0}^{j_1}1 + \sum_{j_1=1}^{d-1}\sum_{j_2=1}^{j_1}\sum_{j_3=1}^{j_2}2 + \sum_{j_1=2}^{d-1}\sum_{j_2=2}^{j_1}\sum_{j_3=2}^{j_2}\sum_{j_4=2}^{j_3}3 +\dots+ \sum_{j_1=n-2}^{d-1}\sum_{j_2=n-2}^{j_1}\sum_{j_3=n-2}^{j_2}\dots\sum_{j_n=n-2}^{j_{n-1}}(n-1)\nonumber\\
&=\sum_{0\leq j_2\leq j_1\leq {d-1}}1+\sum_{1\leq j_3\leq j_2\leq j_1\leq {d-1}}2 + \sum_{2\leq j_4\leq j_3\leq j_2\leq j_1\leq {d-1}}3 +\dots+ \sum_{n-2\leq j_n\leq j_{n-1}\leq \dots \leq j_2\leq j_1\leq {d-1}}(n-1)\nonumber\\
&= \sum_{q=1}^{n-1} \; \bigg[ \,q\,\sum_{q-1\leq j_{q+1}\leq j_q\leq \dots \leq j_1\leq {d-1}}1 \bigg] = \sum_{q=1}^{n-1} \,q\,\binom{d+1}{d-q} = \sum_{q=1}^{n-1} \,q\,\binom{d+1}{q+1}.
\end{align}
\end{widetext}
In the last step of this calculation we used that $\binom{a+b-1}{b-1}$ is the number of distinct ways to draw $a$ times from $b$ options with replacement where the order does not matter, such that with $a=q+1$ and $b=d-1-(q-1)+1$ we get the number of terms in each sum. Inserting this expression into \eqref{eq:degfct} we obtain
\begin{equation}\label{eq:degfinal}
\varrho(L,n,d) = \sum_{q=0}^{n-1} L\binom{d}{q} - q\binom{d+1}{q+1},
\end{equation}
which agrees with Eq.~\eqref{eq:degdist}. We tested \eqref{eq:degdist} for all combinations of locality parameters $n$ and $d$ for systems of sizes up to $L=10$, and also for system sizes up to $L=16$ for smaller values of $n$ and $d$.

\section{Construction of Hamiltonians}
\label{app:constructionH}

The Hamiltonian \eqref{eq:H} can be written as
\begin{align}
H = \sum_{\chi(n)\,|\,(\chi_n-\chi_1\leq d)} h_{\chi_1\chi_2\ldots\chi_n}.
\end{align}
with
\begin{multline}\label{e:h}
\!\!\!\!\!h_{\chi_1\chi_2\ldots\chi_n} = \!\!\!\sum_{\phi_1,\phi_2,\dots,\phi_n}^{(x,y,z)}\!\!\! a_{\chi_1\dots\chi_n}^{\phi_1\dots\phi_n}\1_{2^{\chi_i-1}} \otimes\s_{\chi_1}^{\phi_1} \otimes \1_{2^{\chi_2-\chi_1-1}}\\
\otimes \s_{\chi_2}^{\phi_2} \otimes \dots \otimes \1_{2^{\chi_{n}-\chi_{n-1}-1}} \otimes \s_{\chi_n}^{\phi_n} \otimes \1_{2^{L-\chi_n-1}}.
\end{multline}
In {\tt python}, we can make use of the package {\tt scipy.sparse}. We first create sparse matrices for $\s^{x}$, $\s^{y}$, and $\s^{z}$ in the $\s^z$-eigenbasis, using the coordinate format {\tt coo}. The Kronecker products $\otimes$ in \eqref{e:h} are available as {\tt scipy.sparse.kron}, for efficiency reasons the option {\tt formate='coo'} should again be used. Using Knuth's ``algorithm L'', all permutations of a sequence with $n$ $1$'s and $L-n$ $0$'s are efficiently created and each permutation is used to construct $h_{\chi_1\chi_2\ldots\chi_n}$ \eqref{e:h}. To create all tuples $\phi(n)$, the function {\tt product} from {\tt itertools} is used. Summing over all thus generated terms and inserting the random numbers $a_{\chi_1\dots\chi_n}^{\phi_1\dots\phi_n}$ yields the full Hamiltonian from the EXH ensemble defined in Sec.~\ref{ssec:ens}. On the personal computer we used, this procedure works for system sizes up to $L=13$.

By specifying the data type of the sparse matrices as boolean, {\tt dtype='bool'}, and omitting the random variable $a_{\chi_1\dots\chi_n}^{\phi_1\dots\phi_n}$, mask matrices can be generated efficiently for up to $L=18$ spins on a standard computer. The mask matrices are used as adjacency matrices for the EXA ensemble defined in Sec.~\ref{ssec:ens}.

\begin{figure}[t]\centering
	\includegraphics[trim=2cm 16cm 6cm 2.5cm,clip,width=0.9\linewidth]{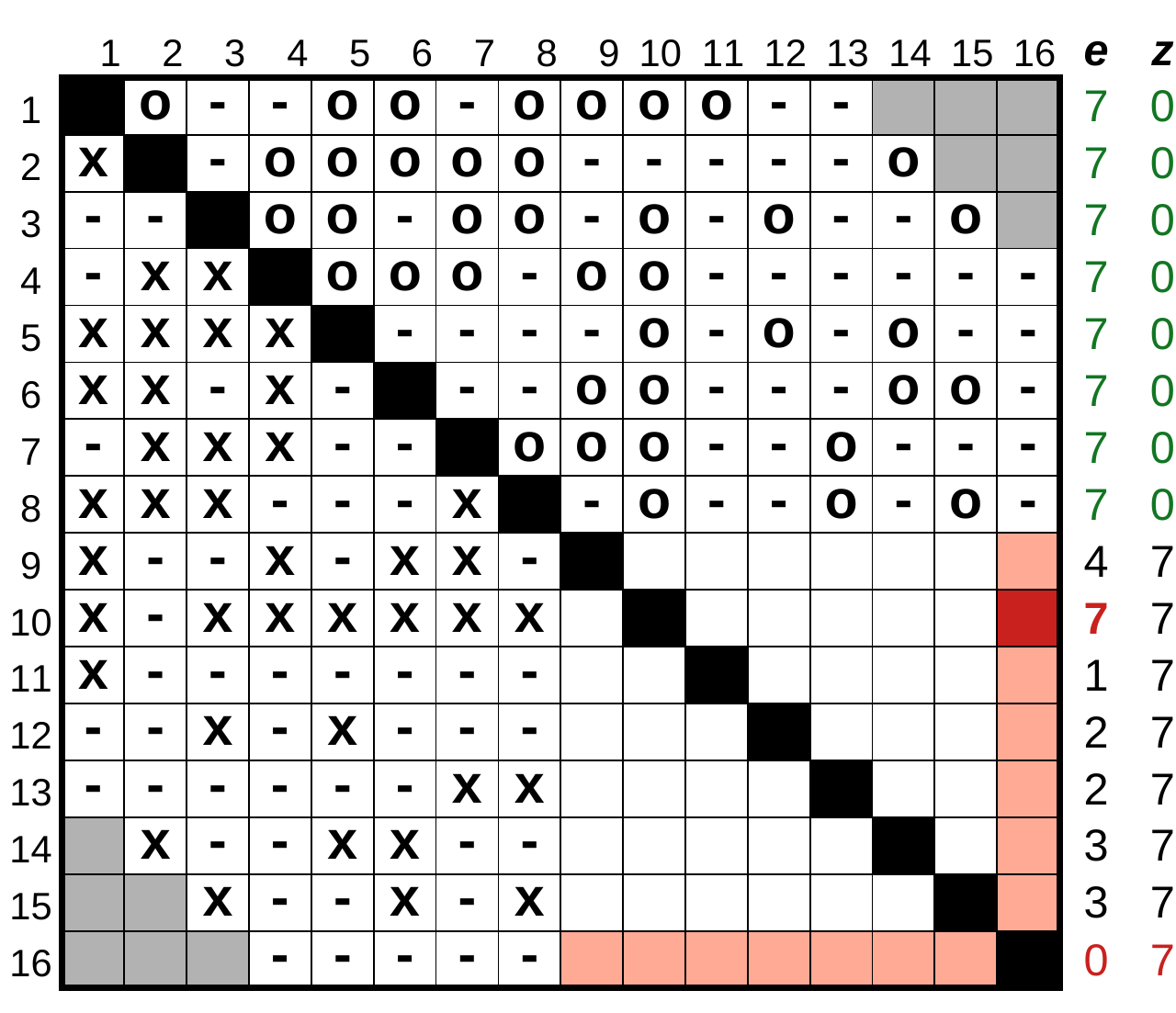}
	\caption{An illustration of a conflict in the progressive construction of regular graphs with banded adjacency matrices for $L=4$. The open circles in the upper triangle show positions of randomly placed entries inside the bandwidth, entries outside the band are shaded in grey. Crosses in the lower triangle mark positions of entries that are required in order to create a symmetric matrix. The two columns on the right show the number of entries ($e$) and the number of empty places ($z$) for each row. Here, the constant degree is $\varrho=7$, so each row will have to have $7$ nonzero entries. The described routine goes through all rows from top to bottom, randomly selects entries in the upper triangle and adds entries in the lower triangle to ensure a symmetric matrix. After completing the $8$th row, the last row has no entries yet and $7$ empty places left and should therefore be completed first before proceeding with row $9$. However, the $10$th row is complete already, and hence either symmetry or constant degree has to be sacrificed. To resolve this conflict, one of the rows $1,4,5,6,7,8$ will be shuffled and other entries be updated accordingly until the $10$ row has an empty place such that the last row can be completed. \label{fig:band_conflict}}
\end{figure}

\section{Random graph ensembles}
\label{app:construction}

\subsection{Construction of adjacency matrices}
\label{app:constructionA}
For the random graph ensembles BRF, BVF, BRC, and REG introduced in Sec.~\ref{ssec:ens}, we need to construct random adjacency matrices with a given constant degree $\varrho$. For the full (non-banded) regular graphs in REG, the function {\tt random\_regular\_graph} in the python package {\tt NetworkX} generates such matrices for a given degree $\varrho$ and number of nodes $N$ uniformly from all possible matrices. We are not aware of similar ready-to-use packages for banded random matrices of a prescribed bandshape. Here we describe a simple, non-optimised algorithm to generate banded random matrices with constant degree. Note that, as opposed to {\tt random\_regular\_graph}, our algorithm does not ensure a uniform sampling of matrices.

\begin{figure}[b!]\centering
	\includegraphics[width=0.8\linewidth]{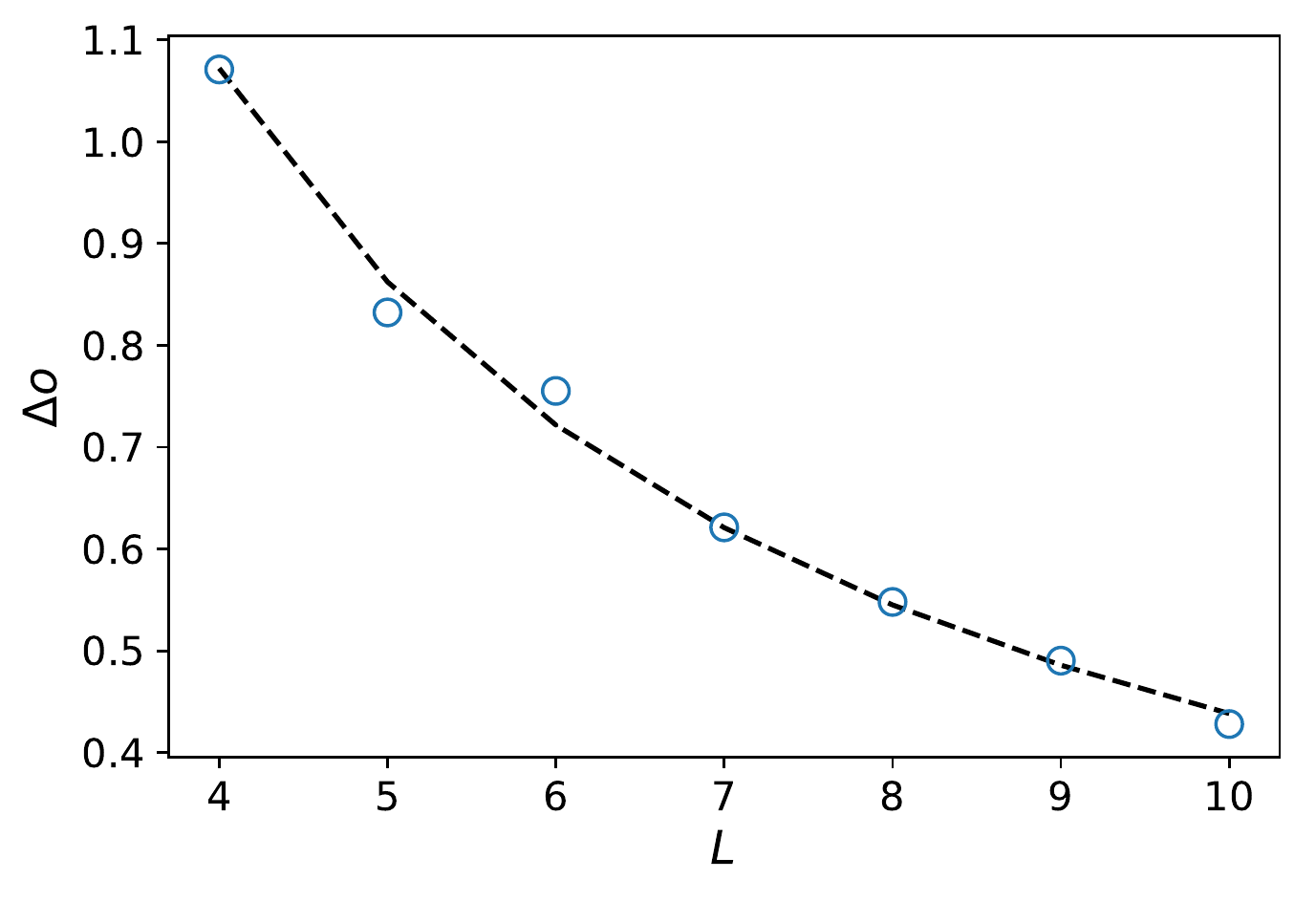}\\	\includegraphics[width=0.8\linewidth]{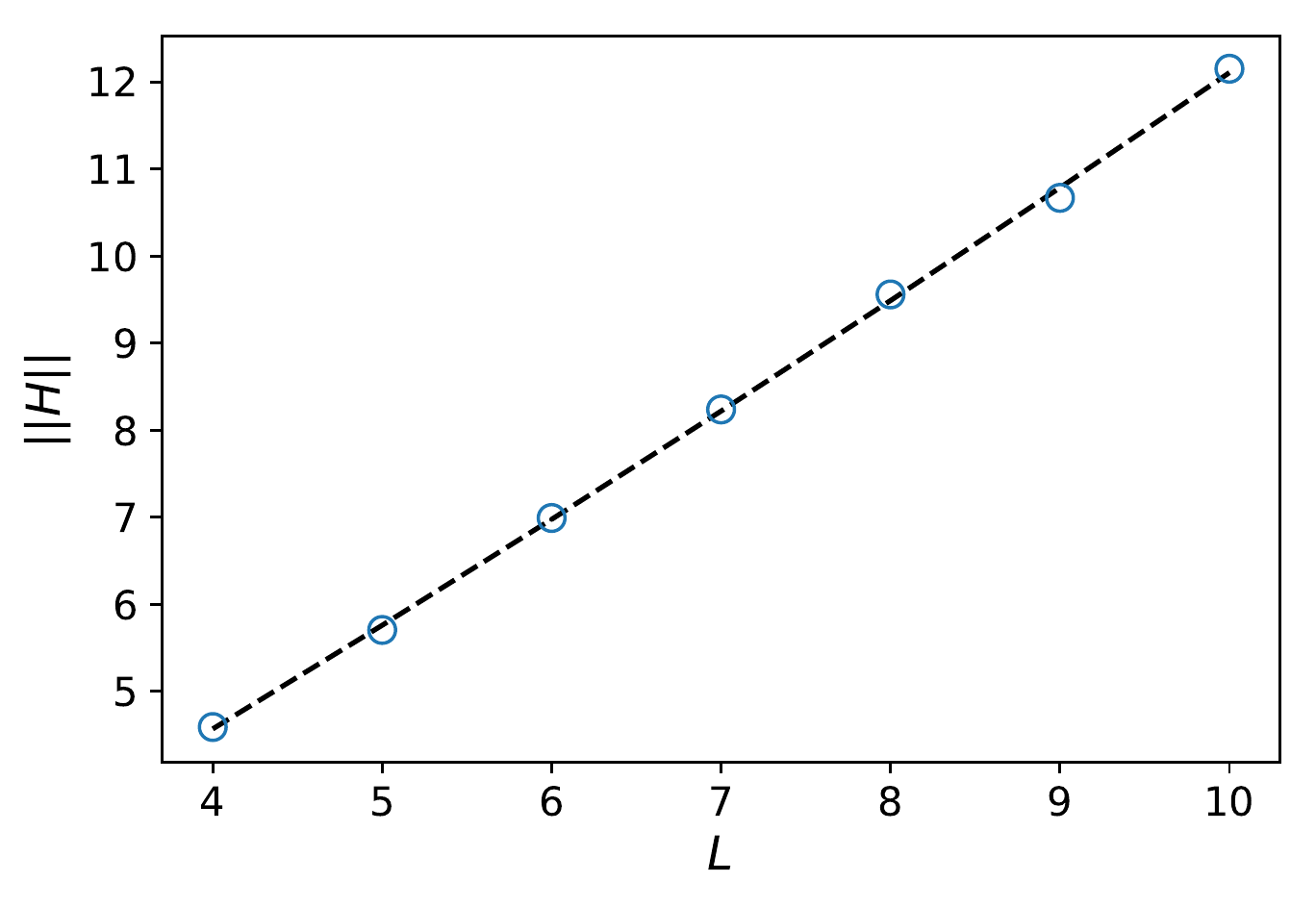}
	\caption{Exponential functions $\exp(aL^b)$ with fit parameters $a$ and $b$, fitted to disorder averages of $\D o$ \eqref{eq:bj} (top) and $\|H\|$ (bottom) for 100 disorder samples. Both plots are for the Hamiltonian \eqref{eq:H} with parameters $n=1$, $d=1$, and system size $L=8$. \label{fig:Do_normH_fit}}
\end{figure}

The challenge is that the adjacency matrices are required to be symmetric. For that reason, going through all rows from top to bottom and randomly choosing a fixed number of nonzero elements within a prescribed bandwidth cannot work, as the corresponding columns must be filled concurrently, which in turn means that rows further down are partly filled as we go. One therefore must keep track of the number $e$ of nonzero elements already placed in each uncompleted row, and choose only as many nonzero entries as required to obtain the constant degree for that row. Moreover, the zero entries of each completed row must remain zero in the corresponding columns, reducing the number $z$ of entries in the uncompleted rows that can be nonzero. Before completing each row with the missing number of nonzero elements, all later rows must be checked for available space. Once a row has just enough space left as needed to obtain the desired constant degree for that row, it must be completed and the number of nonzero entries and space for the uncompleted rows must be updated. When the row has been completed and no other rows need to be completed first, the routine resumes with the topmost uncompleted row.

\begin{figure*}\centering
	\includegraphics[width=0.32\linewidth]{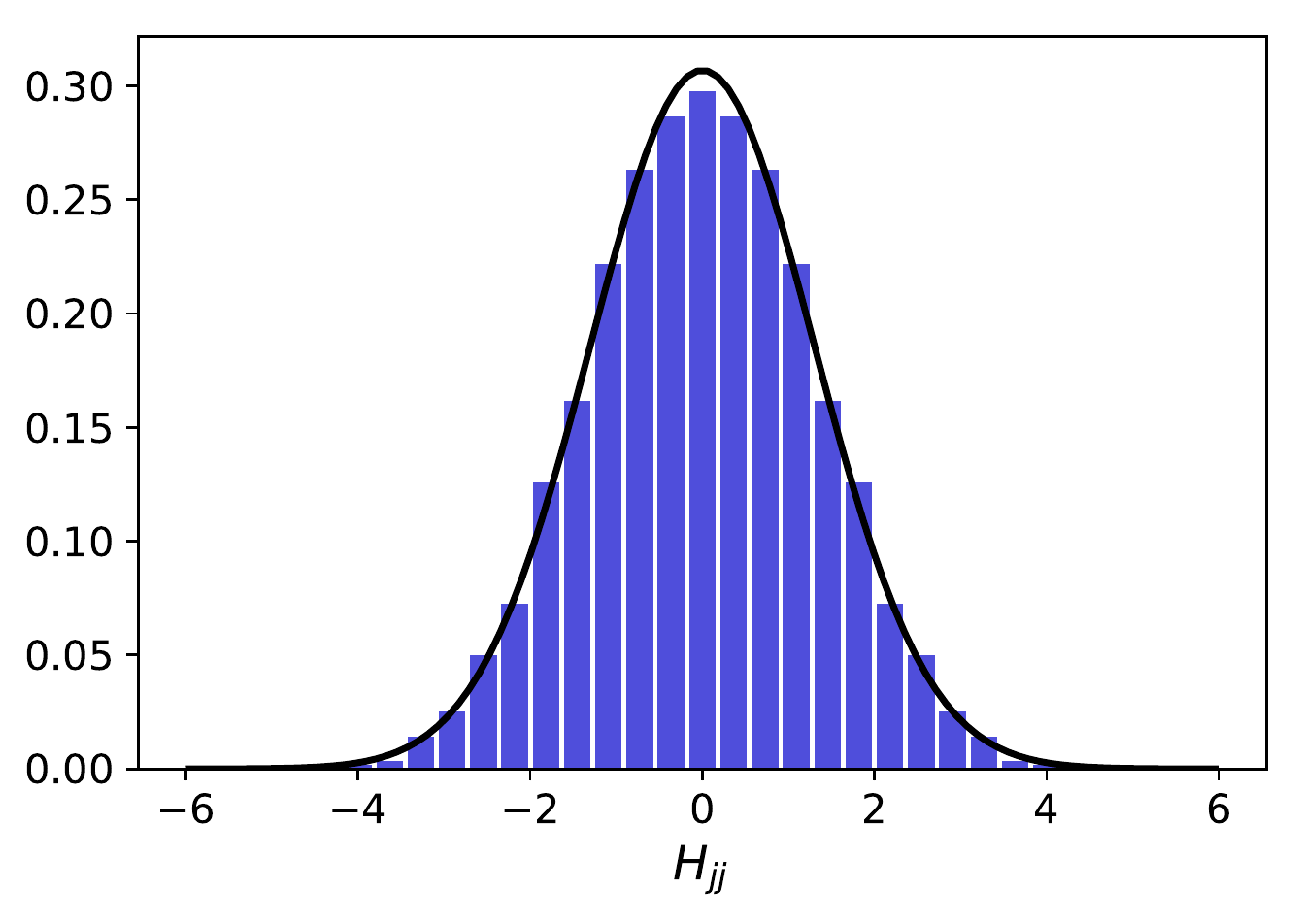}\hfill	\includegraphics[width=0.32\linewidth]{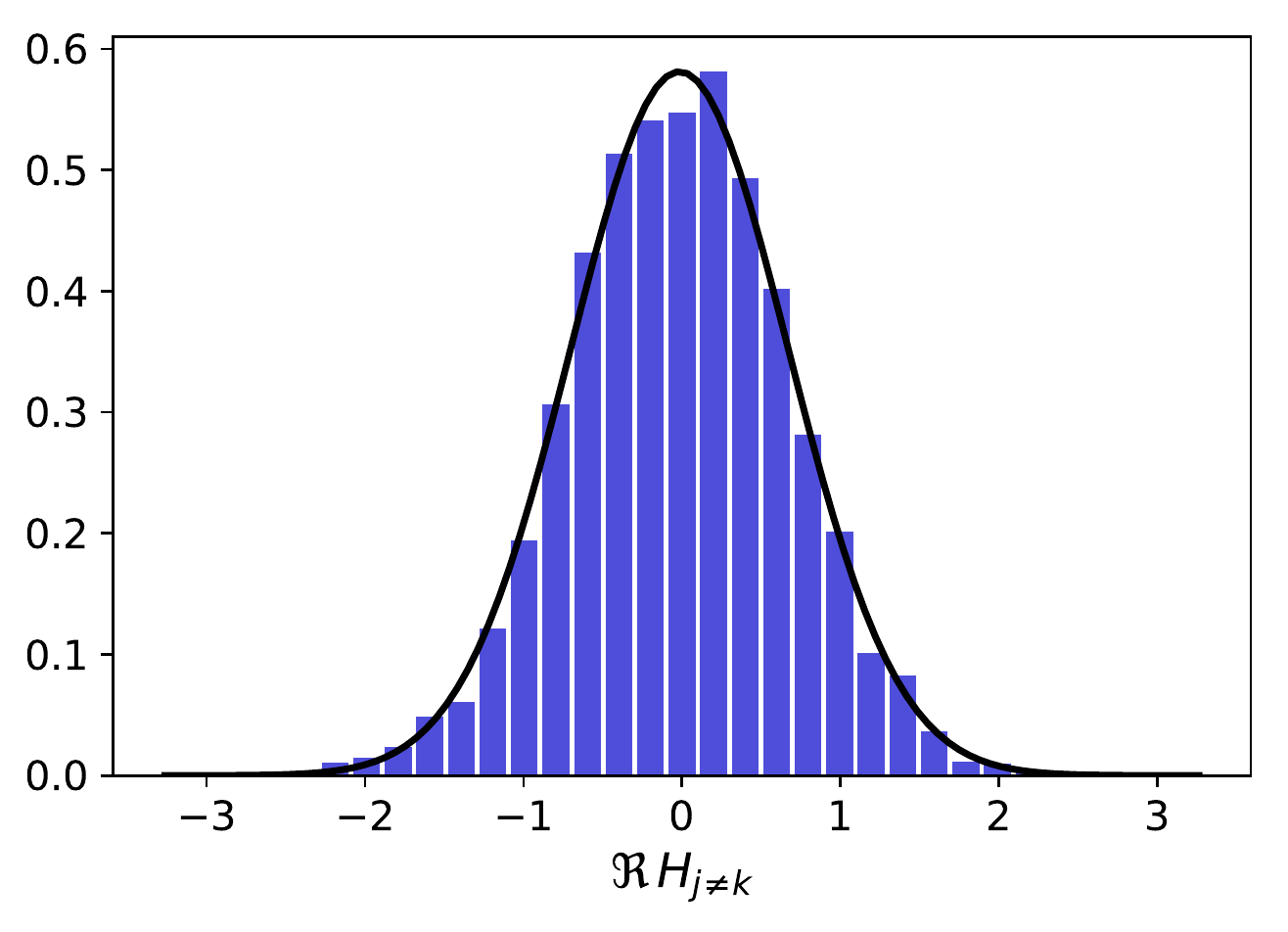}\hfill
	\includegraphics[width=0.32\linewidth]{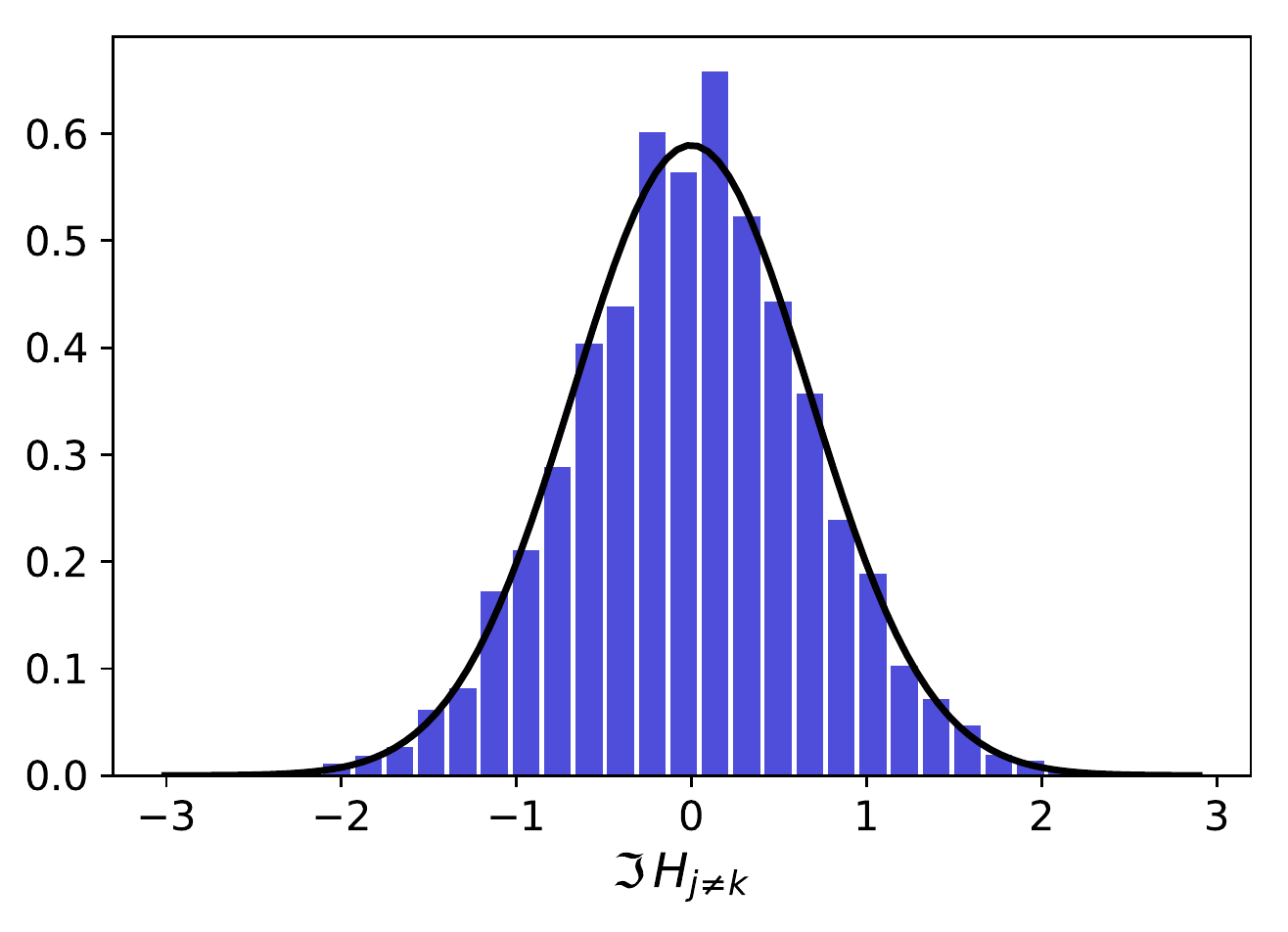}
	\caption{Histograms of diagonal (real-valued) and off-diagonal (real and imaginary parts) matrix elements $H_{jk}$ of the Hamiltonian \eqref{eq:H} with $n=2$ and $d=1$ on $L=8$ lattice sites. Statistics of matrix elements is obtained by generating $100$ samples of Hamiltonians. Mean and standard deviations are estimated and the resulting normal distributions are shown as solid black lines. \label{fig:histdistrs}}
\end{figure*}

The algorithm described so far will most likely run into a conflict: a row with just enough space to obtain the desired degree corresponds to a column that stretches over a row that has already been completed. This situation is illustrated in Fig.~\ref{fig:band_conflict}. The routine could start over until the adjacency matrix is successfully built, but such a strategy is inefficient. Instead, after completion of each row, all uncompleted rows are checked for possible future conflicts. If a conflict is detected, entries in the row that causes the conflict are shuffled randomly until the conflict is resolved. If the conflict cannot be resolved after a predefined number of iterations, the routine starts over. With this strategy, more than $80\%$ of the attempts are successful, and the success rate becomes even higher for larger matrix sizes, which we deem acceptable for our purposes.

With this routine, bands of arbitrary shape can be realised, as long as the bandwidth is larger than the desired constant degree. To determine the node-dependent bandwidth $b$ according to Eqs.~\eqref{eq:bj} and \eqref{eq:Do}, we determine $\bropkt{o_j}{H^\dagger OH-O}{o_j}$ for all eigenstates $\kt{o_j}$ of the observable and pick the maximum value. Since this procedure becomes computationally costly, we fit $\Delta o$ as an exponential function of $L$ to data for system sizes up to $L=10$, and then extrapolate to larger system sizes; see Fig.~\ref{fig:Do_normH_fit} for an illustration. 

\begin{figure}[b!]\centering
\includegraphics[width=0.8\linewidth]{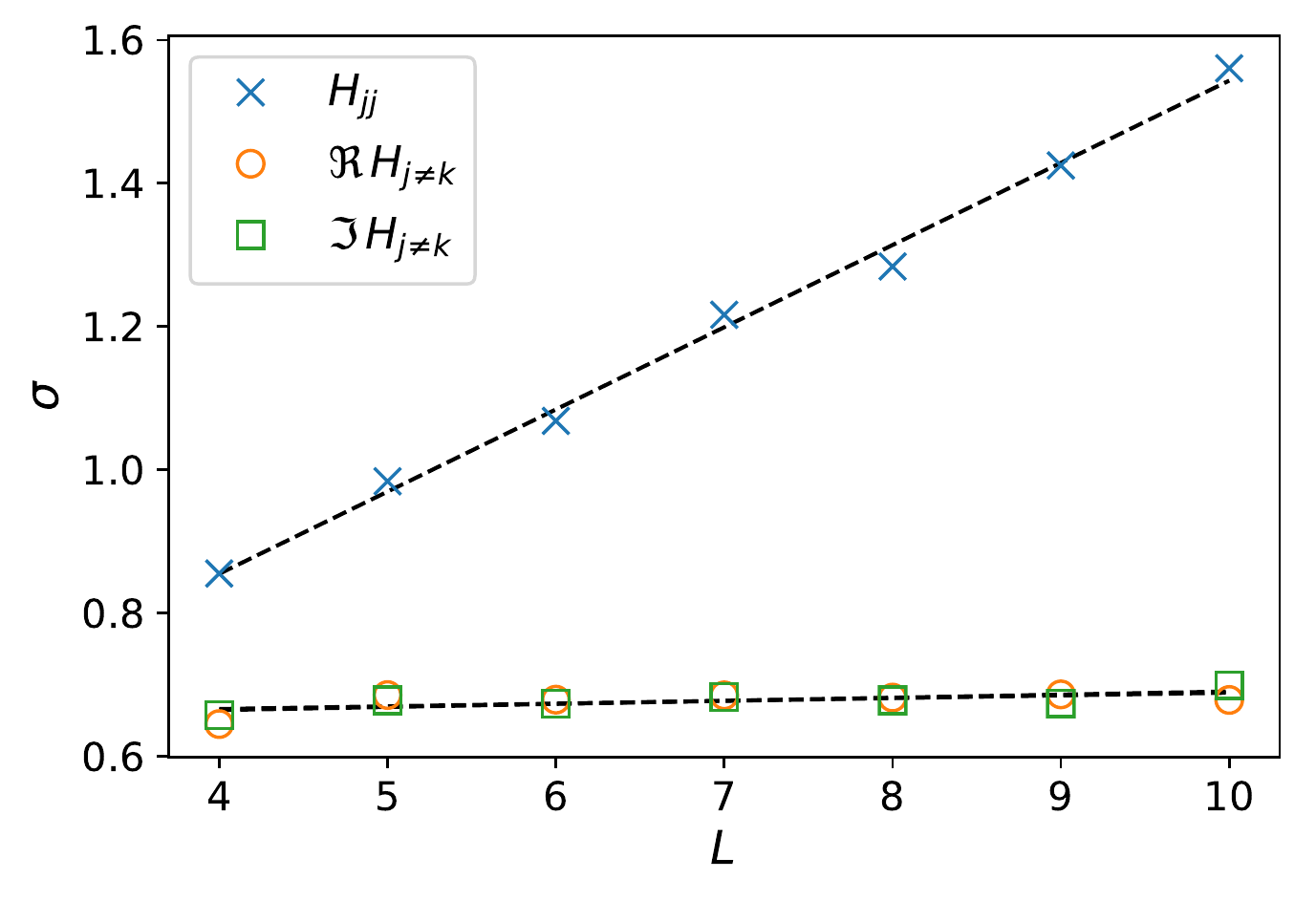}
	\caption{Linear fits of the standard deviations of the distributions of matrix elements $H_{jk}$ shown in Fig.~\ref{fig:histdistrs}. \label{fig:statfit}}
\end{figure}

Using $b$ computed according to this description, together with the constant degree \eqref{eq:degdist}, defines the BRF ensemble. In the BRC ensemble, a constant bandwidth $\max_j b_o(j)$ is used for all nodes. The BVF ensemble also uses the exact bandwidth $b$, but conflicts as described above that may occur in the process of creating the adjacency matrices are not resolved. As a subsequent step after completing the last row, excess edges originating from not resolved conflicts are dropped randomly until the final matrix has the same total number of nonzero elements as the original Hamiltonian. As a result, the degree varies along the nodes for graphs of that ensemble, but the average degree agrees with the constant degree of the graphs in the EXH ensemble.

\begin{figure}[hb]\centering
	\includegraphics[width=\linewidth]{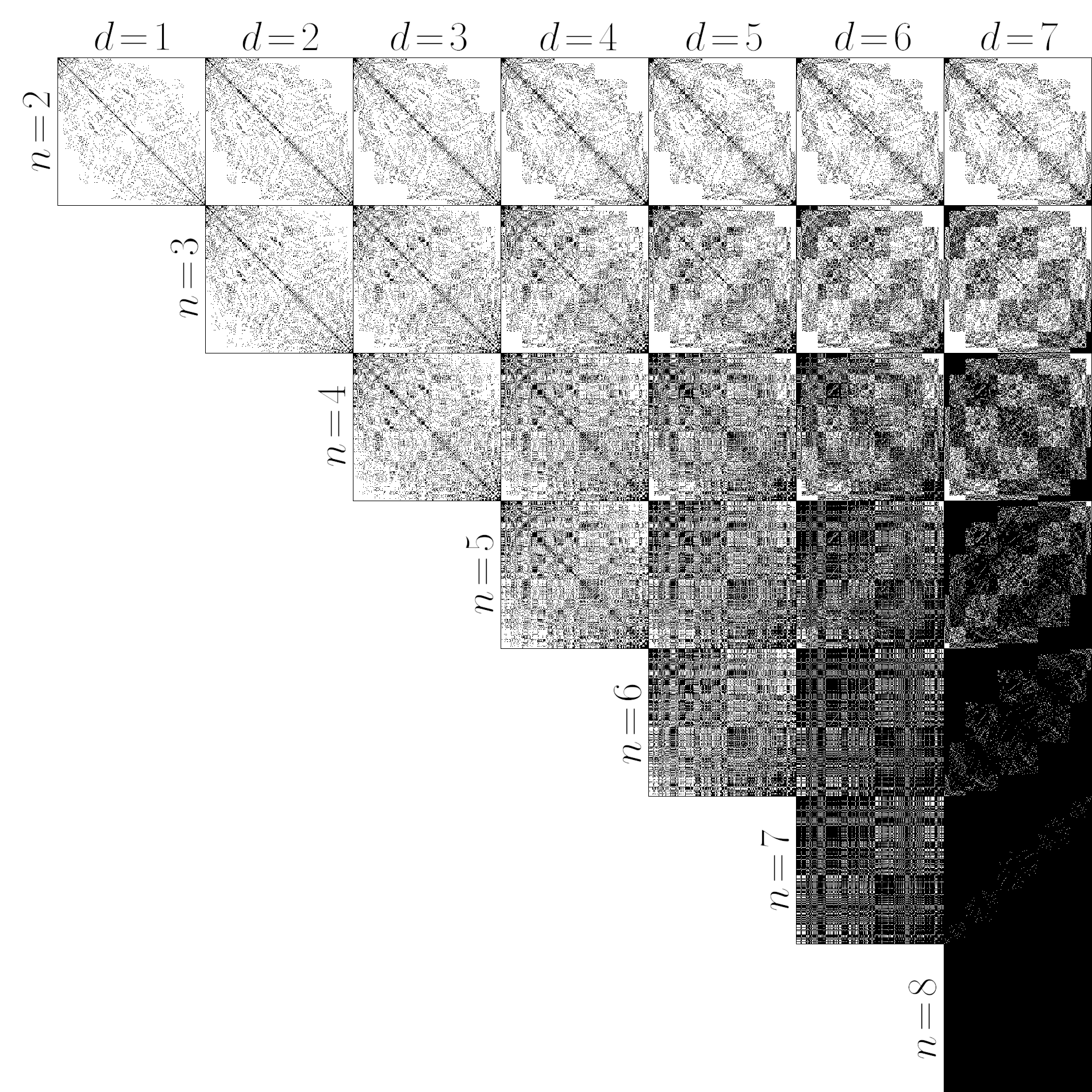}
	\caption{Plots of adjacency matrices $A$ as in Fig.~\ref{fig:masks}, but for magnetisation $M$ with equal weights $a_m(i)\equiv1$.
	The sorting of nodes according to the degenerate eigenvalues of $M$ is not unique. 
	\label{fig:masks_unsrt}}
\end{figure}

\begin{figure*}\centering
\includegraphics[width=0.45\linewidth]{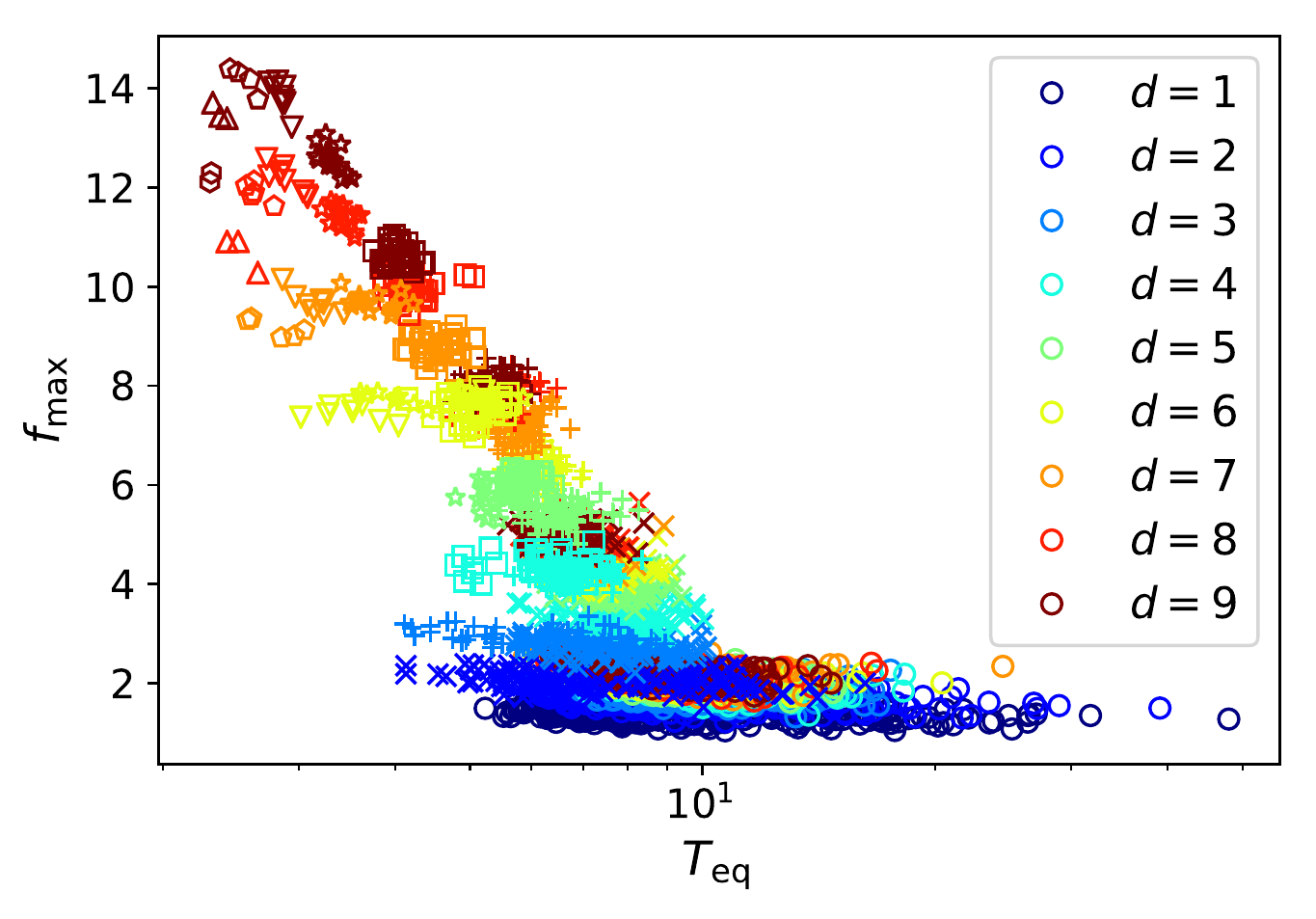}\hfill
\includegraphics[width=0.45\linewidth]{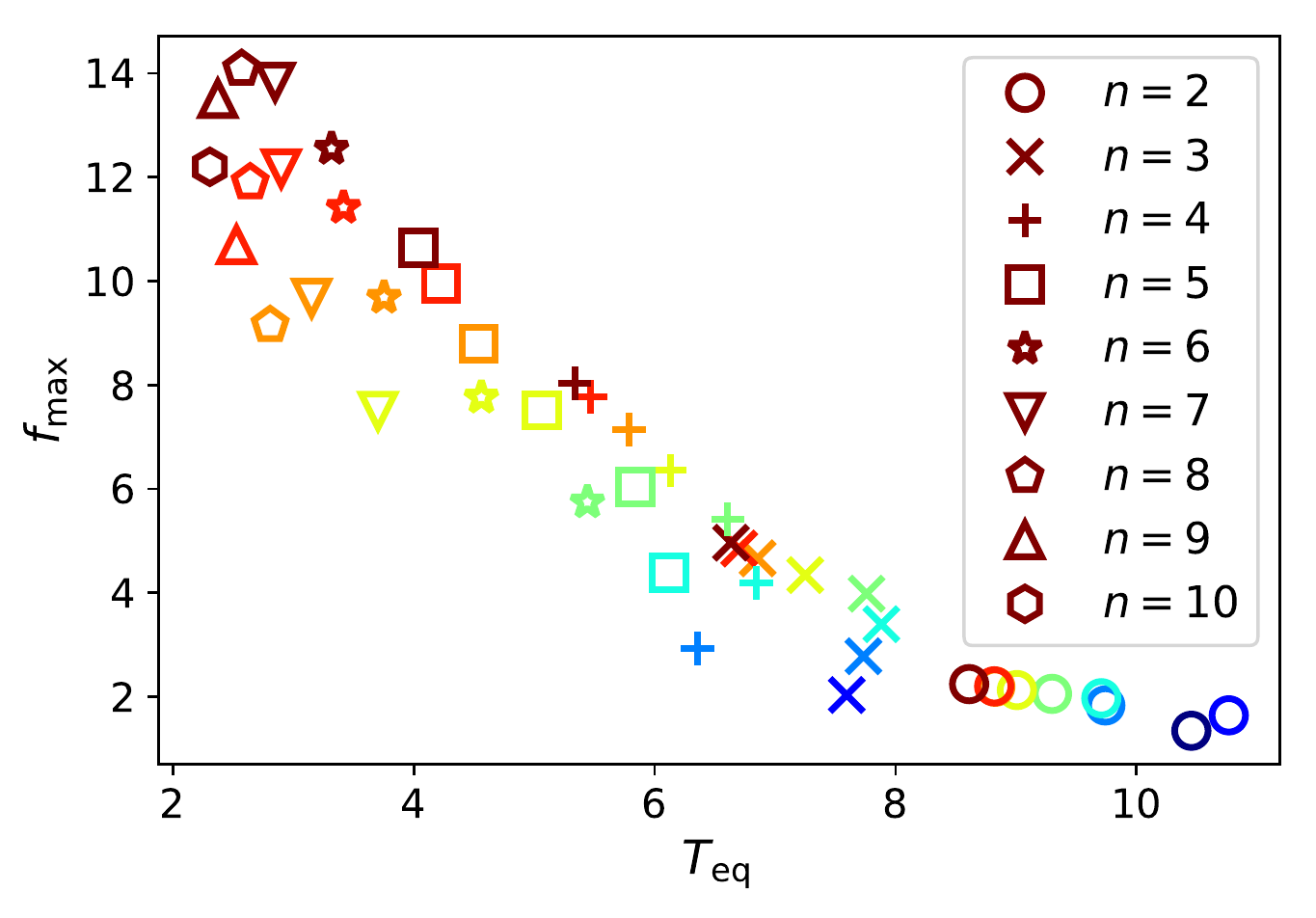}
     \caption{As in Fig.~\ref{fig:Teqs_flowmax}, but using the magnetisation \eqref{eq:M} with equal weights $a_m(i)\equiv1$ as observable. \label{fig:teqs_fmaxs}}
\end{figure*}

\subsection{Weight distributions of random graphs}
\label{app:constructionW}

The adjacency matrices obtained in Sec.~\ref{app:constructionA} characterise the topology of the graph, but not the weights of the edges. The random coupling coefficients of the original Hamiltonian \eqref{eq:H} cause the weights to be random variables which we assume here to be independent and identically distributed (i.i.d.). Since each weight is a sum of many terms involving the coupling coefficients, the central limit theorem can be expected to apply, rendering the weights normally distributed. We verify this expectation by numerically computing the weights $H_{jk}$. The histograms of weights shown in Fig.~\ref{fig:histdistrs} confirm Gaussian weight distributions with zero mean. The standard deviation of the diagonal entries, shown in Fig.~\ref{fig:statfit}, scale linearly with $L$, which ensures extensivity of the Hamiltonian, while the standard deviation of the off-diagonal entries remains constant.

To impose the normalisation $\|H\|=1$ onto the various random graph ensembles we numerically determine $\|H\|$ for several system sizes between 4 and 10. Figure \ref{fig:Do_normH_fit} shows that $\|H\|$ increases linearly with $L$, and we use a fit to these data for extrapolating $\|H\|$ to larger system sizes. For system sizes that are beyond the reach of exact diagonalisation, this procedure is necessary because no algorithms are known that can exploit the sparsity of $H_{jk}$ to efficiently determine spectral properties such as the operator norm.

\section{Non-randomised magnetisation}
\label{app:nonrandom}
In the present paper we mostly used as our observable of choice a randomised version of the magnetisation $M$, with normally distributed random numbers $a_m(i)$ in Eq.~\eqref{eq:M}. This choice is motivated by the convenience of working with an observable that has a nondegenerate spectrum, such that the sorting of eigenstates according to their eigenvalues is unique. In this appendix we use the standard magnetisation with equal weights $a_m(i)\equiv1$. We discuss the difference in the adjacency matrices brought about by this choice of weights, and we demonstrate that the tools and ideas put forward in this paper can nonetheless be applied in this case.

In Fig.~\ref{fig:masks_unsrt} we show the adjacency matrices for all choices of $n$ and $d$ for a system of $L=8$ spins. Unlike in Fig.~\ref{fig:masks}, the bandwidth is block shaped, where each block corresponds to one of the $2L+1$ degenerate eigenvalues of $M$, and the block size is given by the degeneracy of the respective eigenvalue. The order of the nodes corresponding to one eigenvalue is set by the order the eigenstates are determined in \texttt{python}, for which no pattern is apparent. Any node with the eigenvalue closest to zero could work as an equilibrium node. 

To estimate the equilibration times, we calculate the maximum flow value between the two nodes with largest distance to equilibrium ($k=1$ and $k=N$). Since these eigenspaces are nondegenerate, the use of the magnetisation with equal weights does not lead to ambiguities. In Fig.~\ref{fig:teqs_fmaxs} we show the equilibration times $T_\text{eq}$ from the exact time evolution {\em vs.}\ the maximum flow value $f_\text{max}$. The results are similar to those for a randomised magnetisation in Fig.~\ref{fig:Teqs_flowmax}, showing strong anticorrelations between $t_\text{eq}$ and $f_\text{max}$.

\bibliographystyle{unsrtnat}
\bibliography{rmask_refs}

\begin{thebibliography}{56}
\providecommand{\natexlab}[1]{#1}
\providecommand{\url}[1]{\texttt{#1}}
\expandafter\ifx\csname urlstyle\endcsname\relax
  \providecommand{\doi}[1]{doi: #1}\else
  \providecommand{\doi}{doi: \begingroup \urlstyle{rm}\Url}\fi

\bibitem[von Neumann(1929)]{vonNeumann29}
J.~von Neumann.
\newblock Beweis des {E}rgodensatzes und des ${H}$-{T}heorems in der neuen
  {M}echanik.
\newblock \emph{Z. Phys.}, 57:\penalty0 30--70, 1929.
\newblock \doi{10.1007/BF01339852}.

\bibitem[Tasaki(1998)]{Tasaki98}
H.~Tasaki.
\newblock From quantum dynamics to the canonical distribution: General picture
  and a rigorous example.
\newblock \emph{Phys. Rev. Lett.}, 80:\penalty0 1373--1376, Feb 1998.
\newblock \doi{10.1103/PhysRevLett.80.1373}.

\bibitem[Linden et~al.(2009)Linden, Popescu, Short, and Winter]{Linden_etal09}
N.~Linden, S.~Popescu, A.~J. Short, and A.~Winter.
\newblock Quantum mechanical evolution towards thermal equilibrium.
\newblock \emph{Phys. Rev. E}, 79:\penalty0 061103, 2009.
\newblock \doi{10.1103/PhysRevE.79.061103}.

\bibitem[Goldstein et~al.(2010{\natexlab{a}})Goldstein, Lebowitz, Mastrodonato,
  Tumulka, and Zangh\`\i]{Goldstein_etal10}
S.~Goldstein, J.~L. Lebowitz, C.~Mastrodonato, R.~Tumulka, and N.~Zangh\`\i.
\newblock Approach to thermal equilibrium of macroscopic quantum systems.
\newblock \emph{Phys. Rev. E}, 81:\penalty0 011109, 2010{\natexlab{a}}.
\newblock \doi{10.1103/PhysRevE.81.011109}.

\bibitem[Reimann(2010)]{Reimann10}
P.~Reimann.
\newblock Canonical thermalization.
\newblock \emph{New J. Phys.}, 12:\penalty0 055027, 2010.
\newblock \doi{10.1088/1367-2630/12/5/055027}.

\bibitem[Reimann and Kastner(2012)]{KastnerReimann12}
P.~Reimann and M.~Kastner.
\newblock Equilibration of isolated macroscopic quantum systems.
\newblock \emph{New J. Phys.}, 14:\penalty0 043020, 2012.
\newblock \doi{10.1088/1367-2630/14/4/043020}.

\bibitem[Gogolin and Eisert(2016)]{GogolinEisert16}
C.~Gogolin and J.~Eisert.
\newblock Equilibration, thermalisation, and the emergence of statistical
  mechanics in closed quantum systems.
\newblock \emph{Rep. Prog. Phys.}, 79:\penalty0 056001, 2016.
\newblock \doi{10.1088/0034-4885/79/5/056001}.

\bibitem[Mori et~al.(2018)Mori, Ikeda, Kaminishi, and Ueda]{Mori_etal18}
T.~Mori, T.~N. Ikeda, E.~Kaminishi, and M.~Ueda.
\newblock Thermalization and prethermalization in isolated quantum systems: a
  theoretical overview.
\newblock \emph{J. Phys. B}, 51:\penalty0 112001, 2018.
\newblock \doi{10.1088/1361-6455/aabcdf}.

\bibitem[Short and Farrelly(2012)]{ShortFarrelly12}
A.~J. Short and T.~C. Farrelly.
\newblock Quantum equilibration in finite time.
\newblock \emph{New J. Phys.}, 14:\penalty0 013063, 2012.
\newblock \doi{10.1088/1367-2630/14/1/013063}.

\bibitem[Goldstein et~al.(2013)Goldstein, Hara, and
  Tasaki]{GoldsteinHaraTasaki13}
S.~Goldstein, T.~Hara, and H.~Tasaki.
\newblock Time scales in the approach to equilibrium of macroscopic quantum
  systems.
\newblock \emph{Phys. Rev. Lett.}, 111:\penalty0 140401, 2013.
\newblock \doi{10.1103/PhysRevLett.111.140401}.

\bibitem[Goldstein et~al.(2015)Goldstein, Hara, and
  Tasaki]{GoldsteinHaraTasaki14}
S.~Goldstein, T.~Hara, and H.~Tasaki.
\newblock Extremely quick thermalization in a macroscopic quantum system for a
  typical nonequilibrium subspace.
\newblock \emph{New J. Phys.}, 17:\penalty0 045002, 2015.
\newblock \doi{10.1088/1367-2630/17/4/045002}.

\bibitem[Malabarba et~al.()Malabarba, Garc\'{\i}a-Pintos, Linden, Farrelly, and
  Short]{Malabarba_etal14}
A.~S.~L. Malabarba, L.~P. Garc\'{\i}a-Pintos, N.~Linden, T.~C. Farrelly, and
  A.~J. Short.
\newblock Quantum systems equilibrate rapidly for most observables.
\newblock \emph{Phys. Rev. E}, 90:\penalty0 012121.
\newblock \doi{10.1103/PhysRevE.90.012121}.

\bibitem[Farrelly(2016)]{Farrelly2016}
T.~Farrelly.
\newblock Equilibration of quantum gases.
\newblock \emph{New J. Phys.}, 18:\penalty0 073014, 2016.
\newblock \doi{10.1088/1367-2630/18/7/073014}.

\bibitem[Reimann(2016)]{Reimann16}
P.~Reimann.
\newblock Typical fast thermalization processes in closed many-body systems.
\newblock \emph{Nat. Commun.}, 7:\penalty0 10821, 2016.
\newblock \doi{10.1038/ncomms10821}.

\bibitem[Wilming et~al.()Wilming, Goihl, Krumnow, and Eisert]{Wilming_etal}
H.~Wilming, M.~Goihl, C.~Krumnow, and J.~Eisert.
\newblock Towards local equilibration in closed interacting quantum many-body
  systems.
\newblock URL \url{https://arxiv.org/abs/1704.06291}.

\bibitem[Garc\'{\i}a-Pintos et~al.(2017)Garc\'{\i}a-Pintos, Linden, Malabarba,
  Short, and Winter]{GarciaPintos_etal17}
L.~P. Garc\'{\i}a-Pintos, N.~Linden, A.~S.~L. Malabarba, A.~J. Short, and
  A.~Winter.
\newblock Equilibration time scales of physically relevant observables.
\newblock \emph{Phys. Rev. X}, 7:\penalty0 031027, 2017.
\newblock \doi{10.1103/PhysRevX.7.031027}.

\bibitem[{de Oliveira} et~al.(2018){de Oliveira}, Charalambous, Jonathan,
  Lewenstein, and Riera]{deOliveira_etal18}
T.~R. {de Oliveira}, C.~Charalambous, D.~Jonathan, M.~Lewenstein, and A.~Riera.
\newblock Equilibration time scales in closed many-body quantum systems.
\newblock \emph{New J. Phys.}, 20:\penalty0 033032, 2018.
\newblock \doi{10.1088/1367-2630/aab03b}.

\bibitem[Reimann(2019)]{Reimann2019}
P.~Reimann.
\newblock Transportless equilibration in isolated many-body quantum systems.
\newblock \emph{New J. Phys.}, 21:\penalty0 053014, 2019.
\newblock \doi{10.1088/1367-2630/ab1a63}.

\bibitem[Wigner(1955)]{Wigner1955}
E.~P. Wigner.
\newblock Characteristic vectors of bordered matrices with infinite dimensions.
\newblock \emph{Ann. Math.}, 62:\penalty0 548--564, 1955.
\newblock \doi{10.2307/1970079}.

\bibitem[Wigner(1957)]{Wigner1957}
E.~P. Wigner.
\newblock Characteristic vectors of bordered matrices with infinite dimensions
  {II}.
\newblock \emph{Ann. Math.}, 65:\penalty0 203--207, 1957.
\newblock \doi{10.2307/1969956}.

\bibitem[Wigner(1958)]{Wigner1958}
E.~P. Wigner.
\newblock On the distribution of the roots of certain symmetric matrices.
\newblock \emph{Ann. Math.}, 67:\penalty0 325--327, 1958.
\newblock \doi{10.2307/1970008}.

\bibitem[Mehta(2004)]{Mehta2004}
M.~L. Mehta.
\newblock \emph{Random Matrices}.
\newblock Elsevier, Amsterdam, 3rd edition, 2004.

\bibitem[Santos and Torres-Herrera(2017)]{Santos2017}
L.~F. Santos and E.~J. Torres-Herrera.
\newblock Analytical expressions for the evolution of many-body quantum systems
  quenched far from equilibrium.
\newblock \emph{AIP Conf. Proc.}, 1912:\penalty0 020015, 2017.
\newblock \doi{10.1063/1.5016140}.

\bibitem[Torres-Herrera et~al.(2018)Torres-Herrera, Garc{\'{i}}a-Garc{\'{i}}a,
  and Santos]{Torres-Herrera2018}
E.~J. Torres-Herrera, A.~M. Garc{\'{i}}a-Garc{\'{i}}a, and L.~F. Santos.
\newblock Generic dynamical features of quenched interacting quantum systems:
  Survival probability, density imbalance, and out-of-time-ordered correlator.
\newblock \emph{Phys. Rev. B}, 97:\penalty0 060303, 2018.
\newblock \doi{10.1103/PhysRevB.97.060303}.

\bibitem[Santos and Torres-Herrera(2018)]{Santos2018}
L.~F. Santos and E.~J. Torres-Herrera.
\newblock Nonequilibrium many-body quantum dynamics: From full random matrices
  to real systems.
\newblock In F.~Binder, L.~Correa, C.~Gogolin, J.~Anders, and G.~Adesso,
  editors, \emph{Thermodynamics in the Quantum Regime. Fundamental Theories of
  Physics}, pages 457--479. Springer, 2018.
\newblock \doi{10.1007/978-3-319-99046-0_19}.

\bibitem[Torres-Herrera et~al.(2016)Torres-Herrera, Karp, T\'avora, and
  Santos]{Torres-Herrera2016}
E.~J. Torres-Herrera, J.~Karp, M.~T\'avora, and L.~F. Santos.
\newblock Realistic many-body quantum systems vs.\ full random matrices: Static
  and dynamical properties.
\newblock \emph{Entropy}, 18:\penalty0 1--20, 2016.
\newblock \doi{10.3390/e18100359}.

\bibitem[French and Wong(1970)]{French1970}
J.~B. French and S.~S.~M. Wong.
\newblock Validity of random matrix theories for many-particle systems.
\newblock \emph{Phys. Lett. B}, 33:\penalty0 449--452, 1970.
\newblock \doi{10.1016/0370-2693(70)90213-3}.

\bibitem[Bohigas and Flores(1971)]{Bohigas1971}
O.~Bohigas and J.~Flores.
\newblock Two-body random {H}amiltonian and level density.
\newblock \emph{Phys. Lett. B}, 34:\penalty0 261--263, 1971.
\newblock \doi{10.1016/0370-2693(71)90598-3}.

\bibitem[Kota et~al.(2011)Kota, Rela{\~{n}}o, Retamosa, and Vyas]{Kota2011}
V.~K.~B. Kota, A.~Rela{\~{n}}o, J.~Retamosa, and M.~Vyas.
\newblock Thermalization in the two-body random ensemble.
\newblock \emph{J. Stat. Mech.}, 2011:\penalty0 P10028, 2011.
\newblock \doi{10.1088/1742-5468/2011/10/P10028}.

\bibitem[Flambaum and Izrailev(2001)]{FlambaumIzrailev01}
V.~V. Flambaum and F.~M. Izrailev.
\newblock Entropy production and wave packet dynamics in the fock space of
  closed chaotic many-body systems.
\newblock \emph{Phys. Rev. E}, 64:\penalty0 036220, 2001.
\newblock \doi{10.1103/PhysRevE.64.036220}.

\bibitem[Borgonovi et~al.(2019)Borgonovi, Izrailev, and
  Santos]{BorgonoviIzrailevSantos19}
F.~Borgonovi, F.~M. Izrailev, and L.~F. Santos.
\newblock Exponentially fast dynamics of chaotic many-body systems.
\newblock \emph{Phys. Rev. E}, 99:\penalty0 010101, 2019.
\newblock \doi{10.1103/PhysRevE.99.010101}.

\bibitem[Borgonovi and Izrailev(2019)]{BorgonoviIzrailev19}
F.~Borgonovi and F.~M. Izrailev.
\newblock Emergence of correlations in the process of thermalization of
  interacting bosons.
\newblock \emph{Phys. Rev. E}, 99:\penalty0 012115, 2019.
\newblock \doi{10.1103/PhysRevE.99.012115}.

\bibitem[Kota(2001)]{Kota2001}
V.~K.~B. Kota.
\newblock Embedded random matrix ensembles for complexity and chaos in finite
  interacting particle systems.
\newblock \emph{Phys. Rep.}, 347:\penalty0 223--288, 2001.
\newblock \doi{10.1016/S0370-1573(00)00113-7}.

\bibitem[Kota and Chavda(2018)]{Kota2018}
V.~K.~B. Kota and N.~D. Chavda.
\newblock Random $k$-body ensembles for chaos and thermalization in isolated
  systems.
\newblock \emph{Entropy}, 20:\penalty0 1--22, 2018.
\newblock \doi{10.3390/e20070541}.

\bibitem[Vyas and Seligman(2018)]{Vyas2018}
M.~Vyas and T.~H. Seligman.
\newblock Random matrix ensembles for many-body quantum systems.
\newblock \emph{AIP Conf. Proc.}, 1950:\penalty0 030009, 2018.
\newblock \doi{10.1063/1.5031701}.

\bibitem[Seligman et~al.(1985)Seligman, Verbaarschot, and
  Zirnbauer]{Seligman1985}
T.~H. Seligman, J.~J.M. Verbaarschot, and M.~R. Zirnbauer.
\newblock Spectral fluctuation properties of {H}amiltonian systems: The
  transition region between order and chaos.
\newblock \emph{J. Phys. A}, 18:\penalty0 2751--2770, 1985.
\newblock \doi{10.1088/0305-4470/18/14/026}.

\bibitem[Fyodorov et~al.(1996)Fyodorov, Chubykalo, Izrailev, and
  Casati]{Fyodorov1996}
Y.~V. Fyodorov, O.~A. Chubykalo, F.~M. Izrailev, and G.~Casati.
\newblock Wigner random banded matrices with sparse structure: Local spectral
  density of states.
\newblock \emph{Phys. Rev. Lett.}, 76:\penalty0 1603--1606, 1996.
\newblock \doi{10.1103/PhysRevLett.76.1603}.

\bibitem[Erdos and Knowles(2011)]{Erdos2011}
L.~Erdos and A.~Knowles.
\newblock Quantum diffusion and eigenfunction delocalization in a random band
  matrix model.
\newblock \emph{Commun. Math. Phys.}, 303:\penalty0 509--554, 2011.
\newblock \doi{10.1007/s00220-011-1204-2}.

\bibitem[Mirlin et~al.(1996)Mirlin, Fyodorov, Dittes, Quezada, and
  Seligman]{Mirlin1996}
A.~D. Mirlin, Y.~V. Fyodorov, F.-M. Dittes, J.~Quezada, and T.~H. Seligman.
\newblock Transition from localized to extended eigenstates in the ensemble of
  power-law random banded matrices.
\newblock \emph{Phys. Rev. E}, 54:\penalty0 3221--3230, 1996.
\newblock \doi{10.1103/PhysRevE.54.3221}.

\bibitem[Nosov and Khaymovich(2019)]{Nosov2019}
P.~A. Nosov and I.~M. Khaymovich.
\newblock Robustness of delocalization to the inclusion of soft constraints in
  long-range random models.
\newblock \emph{Phys. Rev. B}, 99:\penalty0 224208, 2019.
\newblock \doi{10.1103/PhysRevB.99.224208}.

\bibitem[M{\'{e}}ndez-Berm{\'{u}}dez et~al.(2017)M{\'{e}}ndez-Berm{\'{u}}dez,
  {De Arruda}, Rodrigues, and Moreno]{Mendez-Bermudez2017}
J.~A. M{\'{e}}ndez-Berm{\'{u}}dez, G.~F. {De Arruda}, F.~A. Rodrigues, and
  Y.~Moreno.
\newblock Diluted banded random matrices: Scaling behavior of eigenfunction and
  spectral properties.
\newblock \emph{J. Phys. A}, 50:\penalty0 495205, 2017.
\newblock \doi{10.1088/1751-8121/aa9509}.

\bibitem[Jana and Soshnikov(2017)]{Jana2017}
I.~Jana and A.~Soshnikov.
\newblock Distribution of singular values of random band matrices;
  {M}archenko--{P}astur law and more.
\newblock \emph{J. Stat. Phys.}, 168:\penalty0 964--985, 2017.
\newblock \doi{10.1007/s10955-017-1844-5}.

\bibitem[Bourgade(2018)]{Bourgade2018}
P.~Bourgade.
\newblock Random band matrices.
\newblock \emph{Proc. Int. Cong. Math.}, 3:\penalty0 2745--2770, 2018.
\newblock \doi{10.1142/11060}.

\bibitem[Dumitriu and Zhu(2019)]{Dumitriu2019}
I.~Dumitriu and Y.~Zhu.
\newblock Sparse general {W}igner-type matrices: Local law and eigenvector
  delocalization.
\newblock \emph{J. Math. Phys.}, 60:\penalty0 023301, 2019.
\newblock \doi{10.1063/1.5053613}.

\bibitem[Borgonovi et~al.(2016)Borgonovi, Izrailev, Santos, and
  Zelevinsky]{Borgonovi2016}
F.~Borgonovi, F.~M. Izrailev, L.~F. Santos, and V.~G. Zelevinsky.
\newblock Quantum chaos and thermalization in isolated systems of interacting
  particles.
\newblock \emph{Phys. Rep.}, 626:\penalty0 1--58, 2016.
\newblock \doi{10.1016/j.physrep.2016.02.005}.

\bibitem[Brandino et~al.(2012)Brandino, {De Luca}, Konik, and
  Mussardo]{Brandino2012}
G.~P. Brandino, A.~{De Luca}, R.~M. Konik, and G.~Mussardo.
\newblock Quench dynamics in randomly generated extended quantum models.
\newblock \emph{Phys. Rev. B}, 85:\penalty0 214435, 2012.
\newblock \doi{10.1103/PhysRevB.85.214435}.

\bibitem[Goldstein et~al.(2010{\natexlab{b}})Goldstein, Lebowitz, Tumulka, and
  Zangh{\`i}]{GoldsteinCommentary10}
S.~Goldstein, J.~L. Lebowitz, R.~Tumulka, and N.~Zangh{\`i}.
\newblock Long-time behavior of macroscopic quantum systems.
\newblock \emph{Eur. Phys. J. H}, 35:\penalty0 173--200, 2010{\natexlab{b}}.
\newblock \doi{10.1140/epjh/e2010-00007-7}.

\bibitem[Nickelsen and Kastner(2019)]{Nickelsen2019}
D.~Nickelsen and M.~Kastner.
\newblock Classical {L}ieb-{R}obinson bound for estimating equilibration
  timescales of isolated quantum systems.
\newblock \emph{Phys. Rev. Lett.}, 122:\penalty0 180602, 2019.
\newblock \doi{10.1103/PhysRevLett.122.180602}.

\bibitem[Diestel(2005)]{Diestel2005}
R.~Diestel.
\newblock \emph{Graph Theory}.
\newblock Springer, Berlin, 3rd edition, 2005.

\bibitem[Arad et~al.(2016)Arad, Kuwahara, and Landau]{AradKuwaharaLandau16}
I.~Arad, T.~Kuwahara, and Z.~Landau.
\newblock Connecting global and local energy distributions in quantum spin
  models on a lattice.
\newblock \emph{J. Stat. Mech.}, 2016:\penalty0 033301, 2016.
\newblock \doi{10.1088/1742-5468/2016/03/033301}.

\bibitem[Beugeling et~al.(2015)Beugeling, Moessner, and Haque]{Beugeling2015}
W.~Beugeling, R.~Moessner, and M.~Haque.
\newblock Off-diagonal matrix elements of local operators in many-body quantum
  systems.
\newblock \emph{Phys. Rev. E}, 91:\penalty0 012144, 2015.
\newblock \doi{10.1103/PhysRevE.91.012144}.

\bibitem[Wang and Wang(2017)]{Wang2017}
J.~Wang and W.-G. Wang.
\newblock Correlations in eigenfunctions of quantum chaotic systems with sparse
  {H}amiltonian matrices.
\newblock \emph{Phys. Rev. E}, 96:\penalty0 052221, 2017.
\newblock \doi{10.1103/PhysRevE.96.052221}.

\bibitem[Morampudi and Laumann(2019)]{Morampudi2018}
S.~C. Morampudi and C.~R. Laumann.
\newblock Many-body systems with random spatially local interactions.
\newblock \emph{Phys. Rev. B}, 100:\penalty0 245152, 2019.
\newblock \doi{10.1103/PhysRevB.100.245152}.

\bibitem[Nation and Porras(2018)]{Nation2018}
C.~Nation and D.~Porras.
\newblock Off-diagonal observable elements from random matrix theory:
  distributions, fluctuations, and eigenstate thermalization.
\newblock \emph{New J. Phys.}, 20:\penalty0 103003, 2018.
\newblock \doi{10.1088/1367-2630/aae28f}.

\bibitem[Foini and Kurchan(2019)]{Foini2019}
L.~Foini and J.~Kurchan.
\newblock Eigenstate thermalization hypothesis and out of time order
  correlators.
\newblock \emph{Phys. Rev. E}, 99:\penalty0 42139, 2019.
\newblock \doi{10.1103/PhysRevE.99.042139}.

\bibitem[Brenes et~al.(2020)Brenes, Pappalardi, Goold, and Silva]{Brenes2019}
M.~Brenes, S.~Pappalardi, J.~Goold, and A.~Silva.
\newblock Multipartite entanglement structure in the eigenstate thermalization
  hypothesis.
\newblock \emph{Phys. Rev. Lett.}, 124:\penalty0 040605, 2020.
\newblock \doi{10.1103/PhysRevLett.124.040605}.

\end{thebibliography}

\end{document}